\documentclass[12pt,preprint]{aastex}

\newcommand{\sgra}	{Sgr~A*}
\newcommand{\sgrab}	{Sgr~A*~}
\newcommand{\tnot}	{\ifmmode {\Theta_0}\else {$\Theta_0$} \fi}
\newcommand{\rnot}	{\ifmmode {R_0} \else {$R_0$} \fi}
\newcommand{\onot}	{\ifmmode {\Omega_0}\else {$\Omega_0$} \fi}
\newcommand{\vsun}	{\ifmmode {V_\odot} \else {$V_\odot$} \fi}
\newcommand{\peryr}	{y$^{-1}$}

\newcommand{\perkpc}	{kpc$^{-1}$}
\newcommand{\porm}	{\ifmmode {~\pm~} \else {$~\pm~$} \fi}
\newcommand{\kms}	{\ifmmode {{\rm km~s}^{-1}} \else {km~s$^{-1}$} \fi}
\newcommand{\msun}	{\ifmmode {{\rm M}_\odot} \else  {${\rm M}_\odot$} \fi}
\newcommand{\Msun}	{\ifmmode {{\rm M}_\odot} \else  {${\rm M}_\odot$} \fi}
\newcommand{\lsun}	{\ifmmode {{\rm L}_\odot} \else  {${\rm L}_\odot$} \fi}
\newcommand{\vmax}	{\ifmmode {V_{\rm max}} \else {$V_{\rm max}$} \fi}
\newcommand{\dNdr}	{\ifmmode {dN \over dr} \else {$dN \over dr$} \fi}
\newcommand{\ie}	{i.e.,~}
\newcommand{\eg}	{eg,~}
\newcommand{\etal}	{et al.~}
\newcommand{\fifty}	{{1950}}
\newcommand{\twoth}     {{2000}}
\newcommand{\Mr}        {\ifmmode {M_r} \else {$M_r$} \fi}
\newcommand{\Mtotal}    {\ifmmode {M_{{\rm total}}} \else {$M_{{\rm total}}$} \fi}
\newcommand{\rmin}      {\ifmmode {r_{{\rm min}}} \else {$r_{{\rm min}}$} \fi}
\newcommand{\rmax}      {\ifmmode {r_{{\rm max}}} \else {$r_{{\rm max}}$} \fi}
\newcommand{\rhonot}    {\ifmmode {\rho_0} \else {$\rho_0$} \fi}
\newcommand{\dm}        {\ifmmode {\delta m} \else {$\delta m$} \fi}
\newcommand{\twopi}     {\ifmmode {2\pi}   \else {$2\pi$}   \fi}
\newcommand{\pg}        {{\phantom >}}
\newcommand{\pl}        {{\phantom <}}

\newbox\grsign \setbox\grsign=\hbox{$>$} \newdimen\grdimen \grdimen=\ht\grsign
\newbox\laxbox \newbox\gaxbox
\setbox\gaxbox=\hbox{\raise.5ex\hbox{$>$}\llap
     {\lower.5ex\hbox{$\sim$}}}\ht1=\grdimen\dp1=0pt
\setbox\laxbox=\hbox{\raise.5ex\hbox{$<$}\llap
     {\lower.5ex\hbox{$\sim$}}}\ht2=\grdimen\dp2=0pt

\newcommand{\lax}{\mathrel{\copy\laxbox}}

\shorttitle{The Mass of \sgra}
\shortauthors{Reid \& Brunthaler}

\begin{document}

\title{The Proper Motion of \sgra: II. The Mass of \sgra}

\author{M.~J.~Reid}
\affil{Harvard--Smithsonian Center for Astrophysics, Cambridge, MA 02138}
\email{mreid@cfa.harvard.edu}

\author{A.~Brunthaler}
\affil{Max-Planck-Institut f\"ur Radioastronomie, Auf dem H\"ugel 69, 
 D-53121 Bonn, Germany and Joint Institute for VLBI in Europe, Postbus 2, 
 7990 AA Dwingeloo, The Netherlands}
\email{brunthal@jive.nl}


\begin{abstract}
	We report measurements with the VLBA of the 
position of \sgrab with respect to two extragalactic radio sources
over a period of eight years.
The apparent proper motion of \sgrab relative to J1745--283 is
$6.379\pm0.024$~mas~\peryr\ along a position angle of $209.60 \pm 0.18$ 
degrees, almost entirely in the plane of the Galaxy.  
The effects of the orbit of the Sun around the Galactic center can
account for this motion, and the residual proper motion of \sgrab 
perpendicular to the plane of the Galaxy is $-0.4\pm0.9$~\kms.
A maximum-likelihood analysis of the motion expected for a massive object
within the observed Galactic center stellar cluster
indicates that \sgrab contains more than about 10\% of the 
$\approx4 \times 10^6~\msun$ deduced from stellar orbits.  
The intrinsic size of \sgra, as measured by several investigators, is less
than 1~AU, and the implied mass density of 
$\sim10^{22}$~\msun pc$^{-3}$ is within about three orders of magnitude
of a comparable super-massive black hole within its Schwarzschild radius.
Our observations provide the first direct evidence that a compact radiative 
source at the center of a galaxy contains of order
$10^6$~\msun and provides 
overwhelming evidence that it is in the form of a super-massive black hole.
Finally, the existence of ``intermediate mass'' black holes more massive 
than $\sim10^4$~\msun between roughly $10^3$ and $10^5$~AU from \sgrab are 
excluded.
\end{abstract}


\section{Introduction}

   The case for a super-massive black hole at the center of the
Galaxy is extremely strong.  The proper motions, accelerations, and orbits
of stars about a common gravitational center are now being determined to 
high accuracy at infrared wavelengths \citep{S02,S03,Ghez03}.
A total mass of $\approx4\times10^6$~\msun
\footnote{The amount of mass in the central 100~AU region has been
estimated to be between about 3 and 4 million solar masses, depending on
the method used.  Statistical estimators applied to proper motions give
central masses toward the lower end of this range, 
while fits to stellar orbits favor the higher end of the range \citep{S03,Ghez03}.  
The 3-D motion of IRS~9 favors the higher end of this range \citep{R03},
provided that it is bound to \sgra.
Throughout this paper we adopt a total mass of 4 million solar masses.} 
is required within a radius of about 100~AU.
These dramatic results are fully consistent with the theory that
\sgrab, the compact radio source at the Galactic center, is a super-massive 
black hole.  But must this matter be contained in a super-massive black hole 
(SMBH) and must \sgrab be this SMBH?  

The association of the gravitational mass inferred from 
infrared observations with the
radiative source, \sgra, is supported primarily by two arguments:
1) the very close correspondence (better than $\approx10$~mas or 80 AU in 
projection) between the focal positions of the stellar orbits 
\citep{S02,S03,Ghez03} and the infrared position of \sgra \citep{MREG97,R03}, and 
2) the very short dynamical lifetimes of any massive dark cluster 
that would be required were \sgrab not to contain most of the mass \citep{M98}.
Both arguments, while very strong, are perhaps not yet overwhelmingly so for
two reasons. 
Firstly, \sgrab is exceedingly under-luminous for a SMBH; it's bolometric
luminosity of $\ll 10^{37}$~\lsun \citep{S97} is sub-Eddington for 
even a 1~\msun system.
Thus, for example, a single, radio-loud, X-ray binary could mimic the emission 
from \sgra.  Secondly, the short dynamical cluster lifetime arguments are 
predicated on the evolution of an {\it isolated} dense cluster at the Galactic 
center.   It is possible that a quasi-steady-state condition could occur,
whereby stars that are lost from the cluster (\eg by ``evaporation'') are 
replaced by others falling inward from outside the cluster.  The possibility
of such steady-state conditions has not been addressed.
Regardless of the resolution of these issues, any independent
test of whether or not super-massive black holes exist is very important.

If the compact radio source, \sgra, is indeed the 
gravitational source then it should be nearly at rest at the dynamical 
center of the Galaxy.  \citet{BS99} and \citet{R99} present radio
interferometric data showing that the apparent proper motion of \sgrab,
measured against extragalactic sources, is consistent with that expected 
from the effects of the orbit of the Sun around the Galactic center.  
Removing the effects of the Sun's motion, both
papers conclude that the intrinsic motion of \sgrab in the Galaxy
is less than $\approx20$~\kms.

This paper presents new results on the proper motion of \sgra.
We show that \sgra\ is
indeed nearly stationary at the Galactic center, providing a strong
upper limit to the motion of \sgra\ out of the plane of the Galaxy.
Simple and conservative analyses indicate 
that \sgrab contains at least 10\% of this mass. 
The mass density of \sgra, obtained by combining the lower limit
to the mass directly tied to \sgrab from this paper with the 
apparent size upper limits form VLBI observations 
\citep{Rogers94,Krichbaum98,D01,B04}
is so extreme that the case 
for a super-massive black hole becomes overwhelming.

\section{Observations}

	Our observations using the National Radio Astronomy Observatory's
\footnote{The National Radio Astronomy Observatory is operated by
Associated Universities Inc., under a cooperative agreement with the
National Science Foundation.} Very Long Baseline Array (VLBA) were conducted
between 1995 and 2003.  The results of the observations 
from 1995 through 1997 were reported by Reid \etal (1999, hereafter
Paper I).  The observations and analysis for the data from 1998 through 2003 
were similar to those described in Paper I.
Briefly, the observing sequence involved rapid switching between compact 
extragalactic sources, J1745--283 and J1748--291, and \sgra.
Sources were changed every 15 seconds.  We used \sgrab as the 
{\it phase-reference} source, because it is considerably stronger than 
the background sources and could be detected on individual baselines
with signal-to-noise ratios typically between 10 and 20 in the 7 seconds 
of available on-source time.
We edited and calibrated data using standard tasks in the
Astronomical Image Processing System (AIPS) designed for
VLBA data.  

As discussed in Paper I, the dominant sources of relative position uncertainty
are small errors in the atmospheric model used by the VLBA correlator.  
For some of the epochs analyzed in Paper I, we were able to improve our 
relative position uncertainties by modeling simultaneously the
differenced-phase data for the ``J1745--283 minus \sgra'' and
``J1748--291 minus \sgra'' source pairs and solving for a relative 
position shift for each source pair as well as a single vertical atmospheric 
delay error for each antenna.  
However, this procedure requires high signal-to-noise ratio data
for both background sources for any given baseline within an integration
time of $\sim10$ min.  Unfortunately, J1748--291 weakened
considerably after 1997 and we could not use this calibration technique
for most of our observations, resulting in increased positional uncertainties
for the post 1997 data (except for the 2003 data as discussed below).

For the 2003 observations, we developed an alternative calibration 
approach that involved measuring 
directly the vertical atmospheric delay errors before, during, and after the 
$\approx5.5$~hr tracks on the Galactic center sources.  
Following procedures commonly used for geodetic VLBI observations \citep{W00}, 
we observed about a dozen strong extragalactic radio sources from the
International Celestial Reference Frame catalog \citep{Ma98} in rapid succession 
over a period of about 45~min.
These data were taken at 43~GHz, with eight 8-MHz bands that spanned 470~MHz 
of bandwidth, and residual multi-band delays and fringe rates were calculated
for all sources.  These data were modeled as owing to a vertical atmospheric
delay and delay-rate, as well as a clock offset and clock drift rate, 
at each antenna.
Typical formal uncertainties in the estimates of the vertical atmospheric
delays (expressed as excess path lengths) were 0.2 to 0.3~cm.
Our solutions for atmospheric delay errors include an ionospheric
contribution, expected to be about 0.3~cm of vertical path length, which affects
the phase referenced data and should also be removed when we correct the data.
We did not solve for baseline or Earth's orientation and spin parameters,
as the values used during correlation are sufficiently accurate so as not
to contribute significantly to the atmospheric parameter estimates.  Similarly,
we carefully chose extragalactic sources whose positions are known to better
than 1~mas, so as to avoid position errors contributing significantly to the 
residual delays and rates.  We corrected for the effects of the errors in 
the atmospheric model of the correlator on the phase-referenced data using a 
version of the AIPS task CLCOR provided by L.~Kogan of NRAO.  
The image quality improved, and the 
scatter in the positions of J1745--283, corrected in this manner, 
is consistent with single epoch uncertainties of better than about 
0.1~mas in the each coordinate axis.

For data prior to 1998 the position used during correlation of \sgra, 
the phase-reference source, was inaccurate by $\approx0.12$ arcsec.
For the 1998 and subsequent data, a better position 
for \sgrab was used when correlating the data.  
In order to combine all data, we corrected the pre-1998 data for this
change in position.  This change involved two steps.   The first
order effect is a straightforward translation.
Fitted position offsets were added to the original correlator positions; 
then the new (better) source position was subtracted, yielding 
new position offsets.  A subtle second-order correction is then required.
In the phase-referencing process, residual interferometer phase-shifts
for the phase-reference source (owing to the 0.12 arcsec absolute position 
error for \sgra) are subtracted from the source to be
phase-referenced (\eg J1745--283) and ``interpreted'' as a position
shift {\it based on the position of this source.}  Since, the two 
sources are not at the same position on the sky, this leads to
a second-order correction whose magnitude is roughly
$\Delta\theta_{\rm err} \times \theta_{\rm separation}$.
For our case, $\Delta\theta_{\rm err} = 0.12$~arcsec and
$\theta_{\rm separation} \approx 0.01$~radians, leading
to an expected shift of $\approx1$~mas.  We simulated 
correlator visibility phases for the 0.12~arcsec position error for \sgrab
and fitted them for a position offset for J1745--283.
This indicated that we needed to subtract a second-order position correction 
of ($-0.734,+0.051$)~mas, in the (E,N) directions, from the pre-1998
data for J1745--283.   A similar correction for J1748--291
of ($-1.049,+2.162$)~mas was needed.

	The data in Table~1 summarize our relative position measurements.
We have made no correction for the parallax of \sgra, which would shift
its position by $\le0.1$~mas in each coordinate and would be nearly identical
for all data except the epochs near 1999.8. 
The position offsets are given relative to \sgra, the phase reference source,
even though it is \sgrab that appears to move and not J1745--283 or J1748--291.
Reversing the signs of the offsets, the positions on the sky of \sgra, 
relative to the the strongest background source, 
J1745--283, are plotted in Fig.~1. 
A weighted least-squares fit to the easterly and northerly positions versus time 
(dashed line in Fig.~1) produces a track on the sky that, while close to, 
clearly deviates from the Galactic plane.  This can be explained by the 
known component of the Solar Motion perpendicular to the Galactic plane.

\begin{figure}
\epsscale{0.8}
\plotone{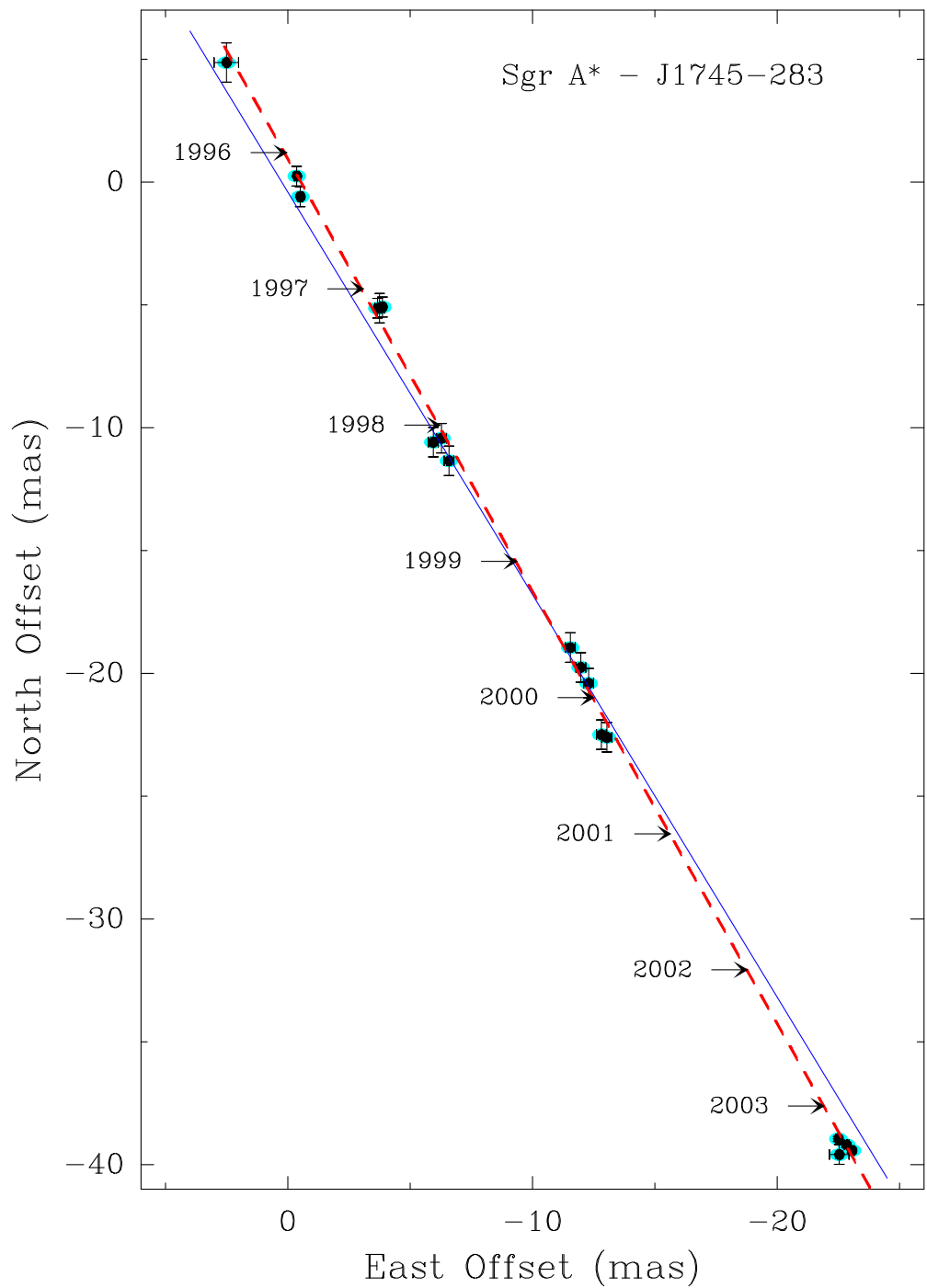}
\caption{\small
Position residuals of \sgrab\ relative to J1745--283 on the plane of the sky.  
Each measurement is indicated with an ellipse, approximating the apparent 
scatter-broadened size of \sgrab at 43 GHz and $1\sigma$ error bars,
which include estimates of systematic uncertainties.
The dashed line is the variance-weighted best-fit proper motion,
and the solid line gives the orientation of the Galactic plane,
which is tilted by $31.40^\circ$ east of north in J2000 coordinates
(see Appendix).
 \label{fig1}}
\end{figure}

The best-fit apparent motion of \sgrab relative to J1745--283
is given in Table~2.  
The uncertainties in Table 2 include estimates of systematic
effects, dominated by small residual errors in modeling atmospheric delays. 
Assuming that J1745--283 is sufficiently distant that
it has negligible intrinsic angular motion, \sgra's apparent motion is 
$-3.151\pm0.018$ and $-5.547\pm0.026$~mas~\peryr\ 
in the easterly and northerly directions, respectively.

\begin{deluxetable}{ccrrrrr}
\tablenum{1}
\tablewidth{0pt}
\tabletypesize{\small}
\tablecaption{Residual Position Offsets Relative to \sgra}
\tablehead{
\colhead{Source}         & \colhead{Date of}      &
\colhead{East Offset}& \colhead{North Offset}&
\colhead{$\ell^{II}$ Offset}    & \colhead{$b^{II}$ Offset} &
 \\
\colhead{}               & \colhead{Observation}   &
\colhead{(mas)}       & \colhead{(mas)    }  &
\colhead{(mas)}       & \colhead{(mas)    }  &
 }
\startdata
J1745--283  & 1995.178 &$  -2.50\porm 0.5$ &$  -4.87\porm 0.8$ &$ -5.46\porm0.63$ &$-0.40\porm0.63$ \\
            & 1996.221 &$   0.37\porm 0.1$ &$  -0.25\porm 0.4$ &$ -0.02\porm0.20$ &$-0.44\porm0.20$ \\
            & 1996.252 &$   0.52\porm 0.1$ &$   0.59\porm 0.4$ &$  0.77\porm0.20$ &$-0.14\porm0.20$ \\
            & 1997.211 &$   3.67\porm 0.1$ &$   5.13\porm 0.4$ &$  6.29\porm0.20$ &$-0.46\porm0.20$ \\
            & 1997.241 &$   3.87\porm 0.1$ &$   5.08\porm 0.4$ &$  6.35\porm0.20$ &$-0.66\porm0.20$ \\
            & 1997.241 &$   3.76\porm 0.2$ &$   5.13\porm 0.6$ &$  6.33\porm0.35$ &$-0.54\porm0.35$ \\
            & 1998.202 &$   5.95\porm 0.2$ &$  10.58\porm 0.6$ &$ 12.13\porm0.35$ &$ 0.43\porm0.35$ \\
            & 1998.219 &$   6.29\porm 0.2$ &$  10.42\porm 0.6$ &$ 12.17\porm0.35$ &$ 0.06\porm0.35$ \\
            & 1998.230 &$   6.59\porm 0.2$ &$  11.34\porm 0.6$ &$ 13.11\porm0.35$ &$ 0.28\porm0.35$ \\
            & 1999.791 &$  11.55\porm 0.2$ &$  18.95\porm 0.6$ &$ 22.19\porm0.35$ &$ 0.01\porm0.35$ \\
            & 1999.799 &$  12.29\porm 0.2$ &$  20.40\porm 0.6$ &$ 23.82\porm0.35$ &$ 0.13\porm0.35$ \\
            & 1999.805 &$  11.97\porm 0.2$ &$  19.75\porm 0.6$ &$ 23.10\porm0.35$ &$ 0.07\porm0.35$ \\
            & 2000.232 &$  13.04\porm 0.2$ &$  22.60\porm 0.6$ &$ 26.08\porm0.35$ &$ 0.64\porm0.35$ \\
            & 2000.238 &$  12.82\porm 0.2$ &$  22.49\porm 0.6$ &$ 25.88\porm0.35$ &$ 0.77\porm0.35$ \\
            & 2003.264 &$  22.51\porm 0.1$ &$  38.94\porm 0.2$ &$ 44.96\porm0.14$ &$ 1.08\porm0.14$ \\
            & 2003.318 &$  22.84\porm 0.1$ &$  39.18\porm 0.2$ &$ 45.34\porm0.14$ &$ 0.92\porm0.14$ \\
            & 2003.339 &$  22.54\porm 0.4$ &$  39.59\porm 0.4$ &$ 45.53\porm0.40$ &$ 1.38\porm0.40$ \\
            & 2003.353 &$  23.06\porm 0.1$ &$  39.41\porm 0.2$ &$ 45.66\porm0.14$ &$ 0.85\porm0.14$ \\
J1748--291  & 1996.221 &$   1.04\porm 0.2$ &$  -2.09\porm 0.6$ &$-1.24\porm0.35$ &$-1.98\porm0.35$ \\
            & 1996.252 &$   1.06\porm 0.2$ &$  -2.18\porm 0.6$ &$-1.31\porm0.35$ &$-2.04\porm0.35$ \\
            & 1997.211 &$   4.53\porm 0.2$ &$   2.76\porm 0.6$ &$ 4.72\porm0.35$ &$-2.43\porm0.35$ \\
            & 1997.241 &$   4.62\porm 0.2$ &$   2.44\porm 0.6$ &$ 4.49\porm0.35$ &$-2.68\porm0.35$ \\
            & 1998.202 &$   7.65\porm 0.2$ &$   8.34\porm 0.6$ &$11.10\porm0.35$ &$-2.19\porm0.35$ \\
            & 1998.230 &$   7.65\porm 0.2$ &$   8.10\porm 0.6$ &$10.90\porm0.35$ &$-2.31\porm0.35$ \\
            & 1999.791 &$  12.87\porm 0.2$ &$  17.79\porm 0.6$ &$21.89\porm0.35$ &$-1.72\porm0.35$ \\
            & 1999.799 &$  12.64\porm 0.2$ &$  17.18\porm 0.6$ &$21.25\porm0.35$ &$-1.84\porm0.35$ \\
            & 1999.805 &$  12.94\porm 0.2$ &$  17.84\porm 0.6$ &$21.97\porm0.35$ &$-1.76\porm0.35$ \\
            & 2000.232 &$  13.98\porm 0.2$ &$  20.05\porm 0.6$ &$24.40\porm0.35$ &$-1.48\porm0.35$ \\
            & 2000.238 &$  14.13\porm 0.2$ &$  20.54\porm 0.6$ &$24.90\porm0.35$ &$-1.36\porm0.35$ \\
            & 2000.246 &$  13.95\porm 0.2$ &$  19.90\porm 0.6$ &$24.25\porm0.35$ &$-1.54\porm0.35$ \\
            & 2003.264 &$  23.42\porm 0.2$ &$  37.89\porm 0.6$ &$44.54\porm0.35$ &$-0.25\porm0.35$ \\
            & 2003.318 &$  23.96\porm 0.2$ &$  37.17\porm 0.6$ &$44.20\porm0.35$ &$-1.09\porm0.35$ \\
\enddata
\tablecomments{Position offsets are relative to \sgra, after removing the
$\approx$0.7 degree differences of the background sources.  
The coordinate offsets are relative to
the following J2000 positions for \sgrab (17 45 40.0409, --29 00 28.118),
J1745--283 (17 45 52.4968, --28 20 26.294), and J1748--291 
(17 48 45.6860, --29 07 39.404).  The conversion to Galactic coordinates
is discussed in the Appendix.  The positions for epochs before 1998
have been corrected for the second-order effects of processing the
phase-reference data from \sgrab with J2000 coordinates of
(17 45 40.0500, --29 00 28.120).
}
\end{deluxetable}

\begin{deluxetable}{cccccll}
\tablenum{2}
\tablewidth{0pt}
\tabletypesize{\small}
\tablecaption{Apparent Relative Motions}
\tablehead{
\colhead{Source -- Reference}         & 
\colhead{Easterly Motion}    & \colhead{Northerly Motion} &
\colhead{$\ell^{II}$ Motion}    & \colhead{$b^{II}$ Motion} &
 \\
\colhead{}               & 
\colhead{(mas y$^{-1}$)}       & \colhead{(mas y$^{-1}$)}  &
\colhead{(mas y$^{-1}$)}       & \colhead{(mas y$^{-1}$)}  &
 }
\startdata
\\
\sgra~--~J1745--283 .......&$-3.151\pm0.018$  &$-5.547\pm0.026$ 
			      &$-6.379\pm0.026$  &$-0.202\pm0.019$ \\
J1748--291~--~J1745--283...&$-0.052\pm0.035$ &$-0.126\pm0.061$ 
			      &$-0.131\pm0.060$ &$-0.021\pm0.037$ \\
\enddata
\tablecomments{Motions are from weighted least-squares fits to the 
data in Table 1.  All results are in the J2000 coordinate system.
Conversion of equatorial to Galactic coordinates in J2000 is discussed
in the Appendix.}
\end{deluxetable}

One potential source of relative positional error is structural variability
of the radio sources.
Since both extragalactic sources and \sgrab could have a core-jet structure,
we could, in principle, be measuring the centroid of a stationary core
and a moving component of an inner-jet.  These effects, however, are likely
quite small as we are observing at a very high frequency, where 
1) stationary cores usually are dominant and
2) time variations are usually rapid ($\ll 8$~y) and will ``average out.''  
Indeed, all of our images are consistent with a point-source, broadened by 
interstellar scattering.  Finally, the results of VLBA observations by 
\citet{BBS01} at 2.3, 5.0 and 8.4~GHz,
when extrapolated to 43~GHz, suggest that source structure is not likely
to significantly affect our relative source positions.

Because we have two reference sources, which should have
independent structures and variations, we can place an observational upper
limit on the magnitude of position wander caused by source structure changes
or any other effects, such as gravitational deflections by intervening stars in
the Galaxy \citep{Hos02}.   
We determined relative positions between the two calibration sources and
plot these in Fig.~2 in the sense 
``J1748--291 minus J1745--283.'' 
The best-fit motions are $-0.052\pm0.035$ and $-0.126\pm0.061$~mas~\peryr\ 
in the easterly and northerly directions, respectively, as indicated 
by the dashed lines.  Were either or both of these Galactic sources,
a much larger relative motion ($>1$~mas~\peryr) would be expected.  
Clearly both are extragalactic.

The {\it uncertainty} in the relative motion of J1748--291 with respect to
J1745--283 is significantly larger than for 
the motion of \sgrab with respect to J1745--283, because 
1) it involves differencing two measured relative positions,
2) the angular separation of the two background sources ($\approx1.0$~deg) 
is greater than between \sgrab and either of the background sources 
($\approx0.7$~deg), which increases sensitivity to systematic errors, and  
3) J1748--291 is the weakest of the three sources and not detected
at all epochs.  
The background sources display little if any motion relative to each other.
The easterly motion of J1748--291 relative to J1745--283 is entirely
consistent with measurement errors, while the measured northerly motion 
is approximately twice the $1\sigma$ error.  Were structural variability
to contribute to our measurement uncertainty, it likely is comparable
to or less than our $1\sigma$ uncertainties.  

\begin{figure}
\epsscale{0.75}
\plotone{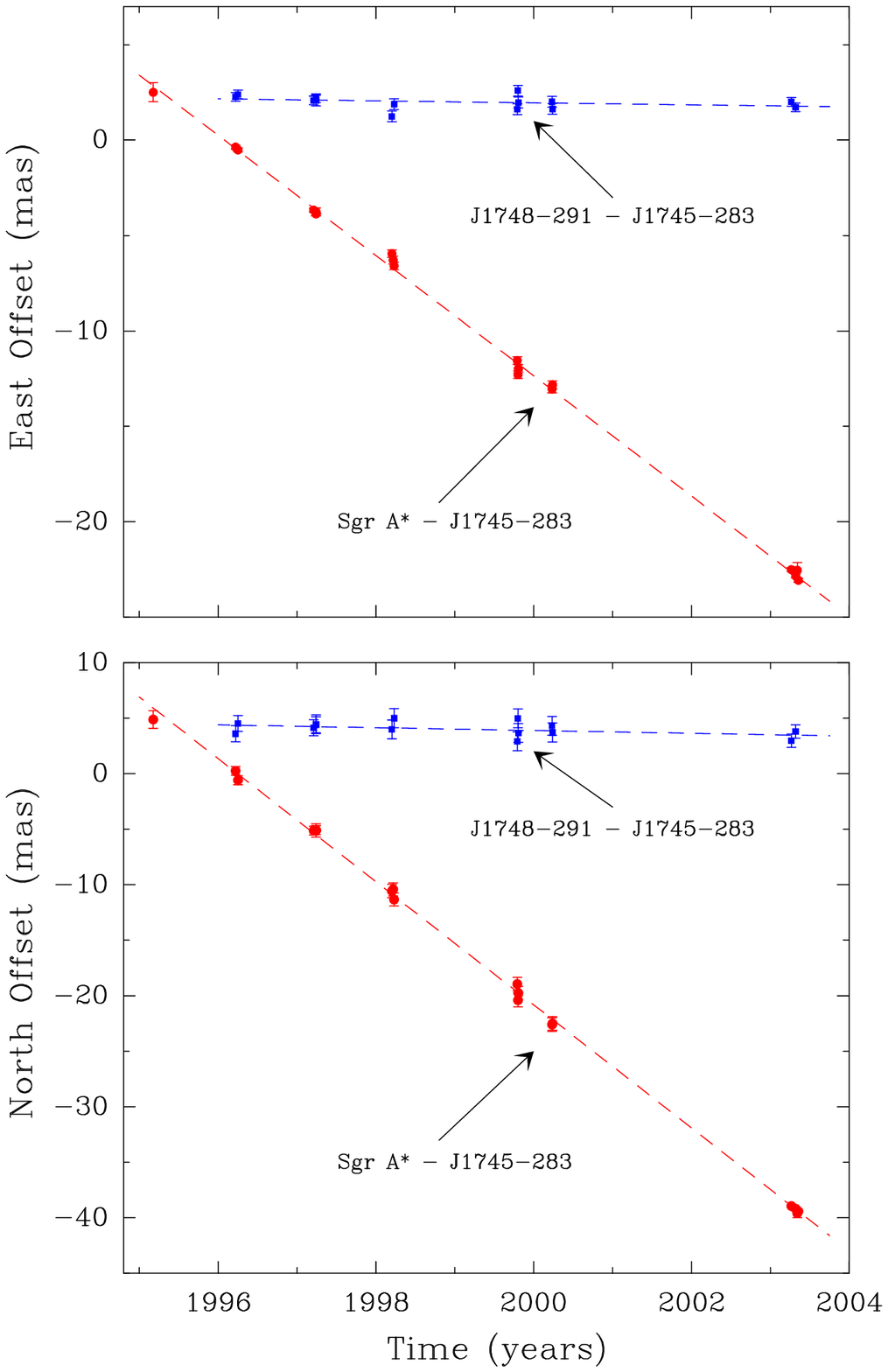}
\caption{\small
Position residuals of \sgrab relative to J1745--283 
and J1748--291 relative to J1745--283  
toward the easterly ({\it top panel}) and northerly ({\it bottom panel})
directions.  The J1748--291 positions are offset for clarity.
Error bars are $\pm1\sigma$ and include estimates of systematic effects.
The position uncertainties are greater for J1748--291, 
owing partly to its weakness and its larger 
angular offset from J1745--283, compared to \sgra. 
The dashed lines are the variance-weighted best-fit proper 
motions.  The position of J1748--291 relative to J1745--283
is constant within $2\sigma$ formal uncertainties and consistent 
with that expected for two extragalactic sources.  
 \label{fig2}}
\end{figure}

\section{Results}

The apparent motion of \sgrab with respect to background
radio sources can be used to estimate the rotation of the Galaxy and
any peculiar motion of the super-massive black hole candidate \sgra.
The apparent motion in the plane of the Galaxy should be dominated by 
the effects of the orbit of the Sun around the Galactic
Center, while the motion out of the plane should contain only small terms
from the Z-component of the Solar Motion and a possible peculiar motion of \sgra.
In the following subsections, we investigate the various components of
the apparent velocity and acceleration of \sgra.

\begin{figure}
\epsscale{0.85}
\plotone{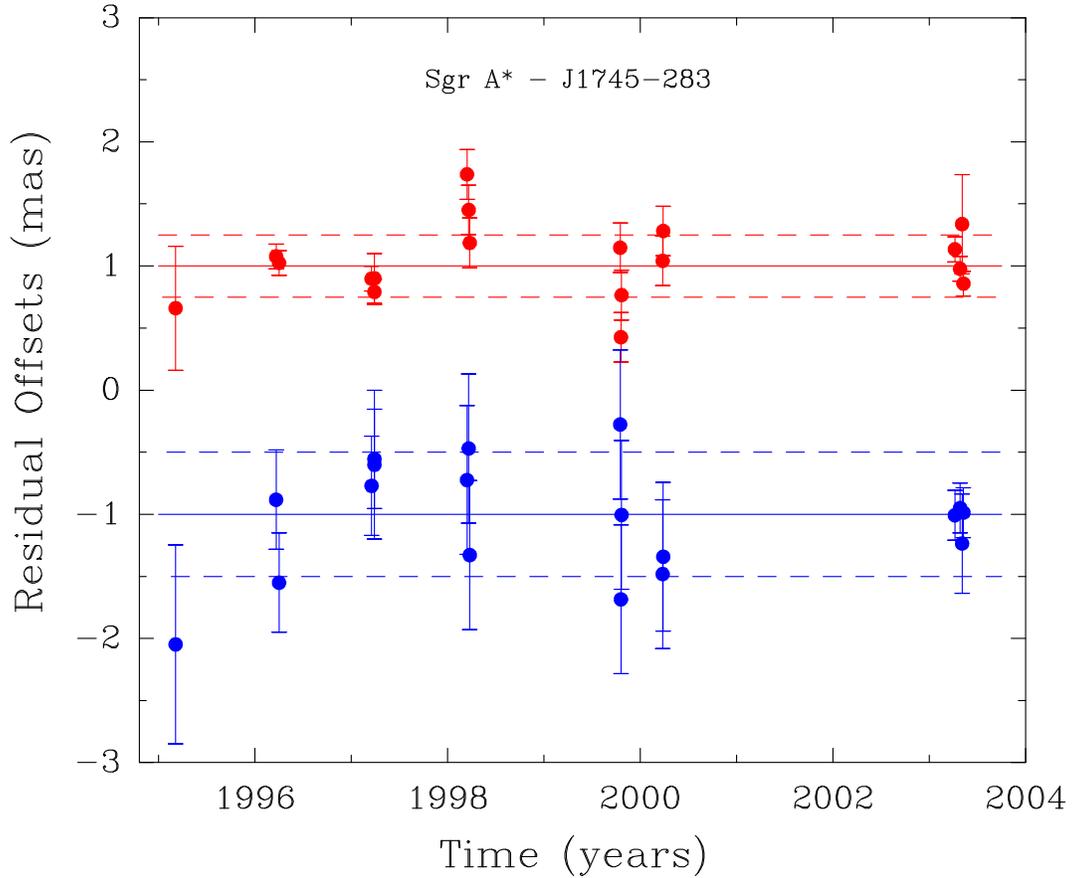}
\caption{\small 
Residual offsets of \sgrab\ relative to J1745--283
with best-fit motions removed.  Easterly residuals are plotted
above northerly residuals, shifted by $+1$ and $-1$~mas, respectively,
for clarity.  
Solid lines indicate zero residual with respect to the best
fitting motions shown in Fig.~2, and dashed lines indicate
estimated limits for any short-period position excursions of \sgra.
Data close in time (\ie, $\approx1$ week) appear slightly correlated,
possibly owing to atmospheric systematics.
Error bars are $1\sigma,$ including
systematic effects of mis-modeled atmospheric delays.  
 \label{fig3}}
\end{figure}

\subsection {Motion of \sgrab in the Plane of the Galaxy}

It is clear from Fig.~1 that the apparent motion of \sgrab is almost 
entirely in the Galactic plane.  Thus, we convert the 
positions from equatorial to Galactic coordinates and determine
motions in Galactic coordinates.  
(Because of the high accuracy of our observations, some pitfalls in the 
implementation of the equatorial to Galactic coordinate conversion (Lane 1979),
and the need to transfer the IAU--defined plane from B1950 to J2000
coordinates, we document the procedures involved in the Appendix.)
Fig.~4 is a plot the position of
\sgrab relative to J1745--283 in Galactic coordinates.  Variance-weighted
least-squares fits of straight lines to these data are indicated by dashed lines.
The apparent motion of \sgrab is
$-6.379\pm0.026$ and $-0.202\pm0.019$~mas~\peryr\ in 
Galactic longitude and latitude, respectively.  

\begin{figure}
\epsscale{1.0}
\plotone{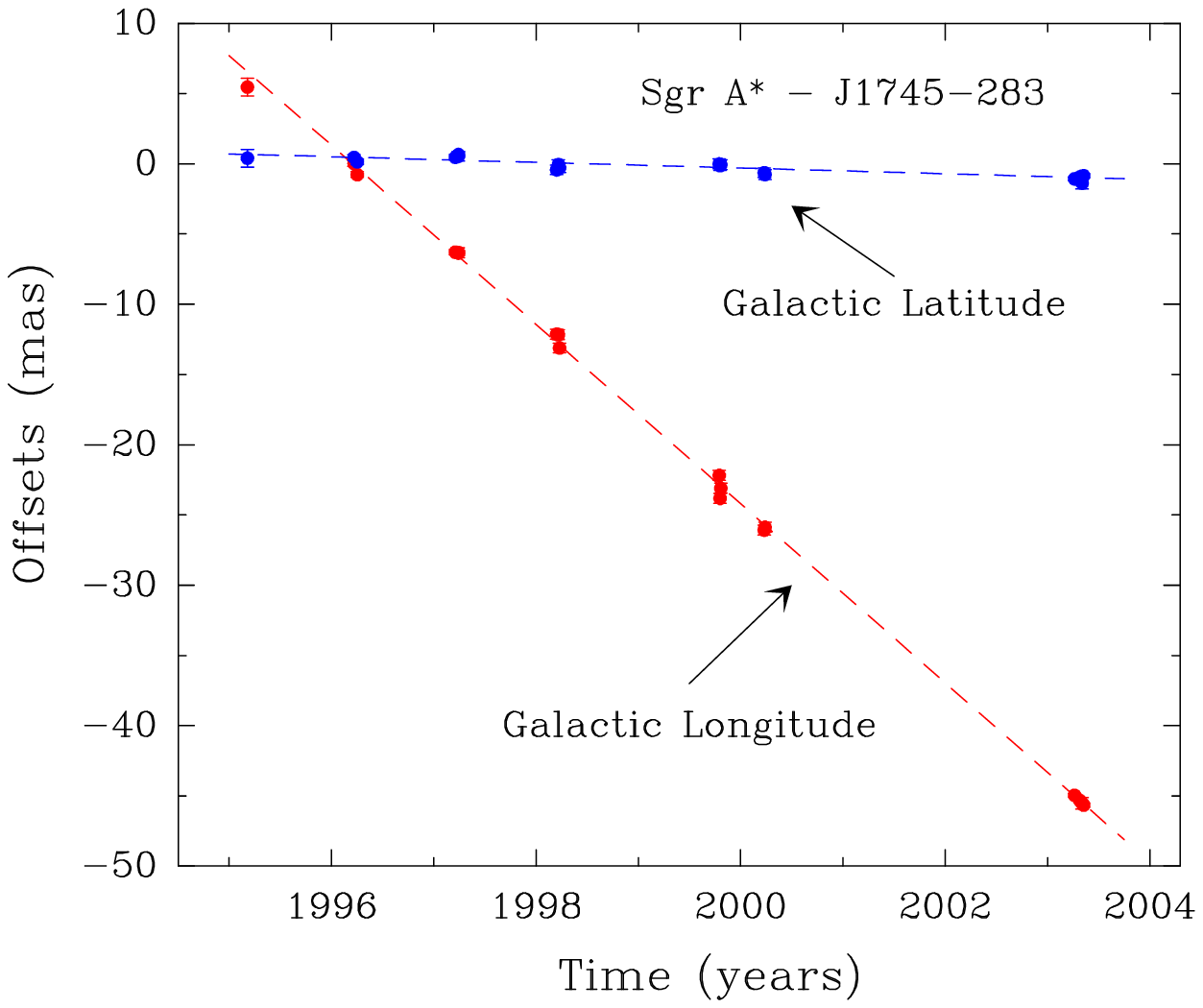}
\caption{\small 
Position residuals of \sgrab\ relative to J1745--283
in Galactic coordinates.  Galactic longitude 
and latitude components are indicated along
with their $1\sigma$ uncertainties.
Dashed lines give the variance-weighted
best-fit components of proper motion.
 \label{fig4}}
\end{figure}

Assuming a distance to the Galactic center ($\rnot$) of $8.0\pm0.5$~kpc 
(Reid 1993), 
the apparent angular motion of \sgrab in the plane of the Galaxy
translates to $-241\pm15$~\kms.  The uncertainty from measurement 
error alone is only $1$~\kms, and the quoted value is dominated
by the 0.5~kpc uncertainty in \rnot.
Provided that the peculiar motion of \sgrab is small (see \S 3.2), 
this corresponds to the reflex of true orbital motion of the Sun around the 
Galactic Center.  This reflex motion can be parameterized
as a combination of a circular orbit (\ie of the LSR)
and the deviation of the Sun from that circular orbit (the Solar Motion).
The Solar Motion, determined from Hipparcos data 
by Dehnen \& Binney (1998), is $5.25\pm0.62$~\kms\ in the direction of 
Galactic rotation.  Removing this component of the Solar Motion from 
the {\it reflex} of the apparent motion of \sgrab yields an estimate for 
\tnot\ of $236\pm15$~{\hbox \kms.}   Note that other definitions and measurements
of this component of the Solar Motion have resulted in somewhat greater
values, \eg 12~\kms \citep{Allen00}, which if adopted would reduce our
value of \tnot to 229~\kms.  
Should \rnot be determined independently 
to high accuracy, then our measurement of the apparent motion of \sgrab would
give \tnot with corresponding accuracy.  
Orbital solutions for stars near \sgrab that combine 
proper motions and radial velocities have great
potential to accomplish this \citep{E03,Ghez03}.

A direct comparison of our measurement 
of the {\it angular} rotation rate of the LSR at the Sun (\tnot/\rnot)
can be made with Hipparcos measurements based on motions of Cepheids.
\citet{FW97} conclude that the angular velocity of circular 
rotation at the Sun, 
\tnot/\rnot (= Oort's A--B), is $27.19 \porm 0.87$~\kms\ \perkpc\ 
($218 \porm 7$~\kms\ for $\rnot=8.0$~kpc).  
Our value of $\tnot/\rnot$, obtained by removing the Solar Motion in
longitude from the {\it reflex} of the motion of \sgrab in longitude,
is $29.45\pm0.15$~\kms\ \perkpc.  The difference between the VLBA and Hipparcos  
angular velocities is $2.26\pm0.9$~\kms\ \perkpc; these
measurements are marginally consistent. 
Neither measurements are sensitive to the value of \rnot, as it is primarily used 
only to remove the small contribution of the Solar Motion.  
Other measurements of $\tnot/\rnot,$ for example from proper motions of 
halo stars relative to galaxies by \citet{Kal04}, yield consistent
values with slightly greater observational uncertainty.

Our value of $\tnot/\rnot$ is a true ``global'' measure of the 
angular rotation rate of the Galaxy,
as opposed to those derived from Oort's constants, which indirectly
determine \tnot from the shear and vorticity in the velocity field of 
material in the solar neighborhood \citep{KL86}.    
The small difference between the local (A--B) and global measures of $\tnot/\rnot$ 
suggests that local variations in Galactic dynamics ($d(\Theta/R)/dR$) are 
less than $\approx3$~\kms~\perkpc.

We now estimate the peculiar motion of \sgrab in the direction of Galactic
rotation by subtracting the Hipparcos-based angular rotation rate of the
Galaxy from the VLBA angular motions of \sgra.
After removing the current best estimate of the motion of the Sun around the
Galactic Center of 223 ($218+5.25$)~\kms \citep{FW97,DB98} from 
our VLBA observation of $241$~\kms, we find
the peculiar motion of \sgrab is $-18\pm7$~\kms\ toward positive Galactic 
longitude (see Table 3).  This estimate of the ``in-plane'' motion of \sgrab comes from 
differencing two {\it angular} motions.  Since this difference is small, 
the uncertainty in \rnot\ does not strongly affect this component of the 
peculiar motion of \sgra.  It is unclear at this time whether or not the
estimate of this component of the peculiar motion of \sgrab differs 
significantly from zero and, if so, if this indicates a difference between
the global and local measures of the angular rotation rate of the Galaxy
or a much larger peculiar motion for \sgrab in the plane of the Galaxy
compared to out of the plane (see \S3.2).
	
Table~3 summarizes the apparent motion of \sgrab in Galactic coordinates with
various known sources of motion removed.  Clearly, these results are of
great interest in regard to the structure and kinematics of the Galaxy
and will be the subject of a later paper.

\begin{deluxetable}{lrrll}
\tablenum{3}
\tablewidth{0pt}
\tabletypesize{\small}
\tablecaption{\sgra's Motion in Galactic Coordinates$^a$}
\tablehead{
\colhead{Description}    & 
\colhead{$\ell^{II}$}    & \colhead{$b^{II}$} &
 \\
\colhead{}               & 
\colhead{(km s$^{-1}$)}       & \colhead{(km s$^{-1}$)}  &
 }
\startdata
Observed \sgrab motion			&$-241\pm15$	&$-7.6\pm0.7$ \\

Effects of Solar Motion$^b$ removed		&$-236\pm15$	&$-0.4\pm0.8$ \\

Effects of Galactic Rotation$^c$ removed	&$-18\pm\phantom{0}7$	&$-0.4\pm0.9$ \\
\enddata

\tablenotetext{a}{Motions are for \sgrab relative to J1745--283 from fitting
results in Table 2.  Speeds assume $\rnot=8.0\porm0.5$ kpc (Reid 1993).  
Quoted uncertainties are $1\sigma$ and include measurement
uncertainty and an angular-to-linear motion scaling error
from the uncertainty in \rnot.}

\tablenotetext{b}{Adopted Solar Motion with respect to a circular orbit
is $5.25\porm0.62$ \kms\ in $\ell^{II}$ and $7.17\porm0.38$ \kms\ 
in $b^{II}$ (Dehnen \& Binney 1998).}

\tablenotetext{c}{Adopted circular rotation of
$27.19\porm0.87$ \kms\ \perkpc\ (Feast \& Whitelock 1997) removed from
our measured angular rotation rate of $-29.45\porm0.15$~\kms~\perkpc\ (after
correction for the Solar Motion) and then multiplied by $\rnot=8.0$~kpc.  
The quoted uncertainty is dominated by measurement uncertainties for the
adopted circular rotation of the Galaxy, scaled by \rnot.
The 0.1~degree uncertainty in the tilt of the Galactic pole \citep{Blaauw60}
causes the slight increase in the uncertainty in the $b$-direction.}

\end{deluxetable}

\subsection	{Motion of \sgrab out of the Plane of the Galaxy}

	Whereas the orbital motion of the Sun (around the Galactic Center)
complicates estimates of the ``in-plane" component of the 
peculiar motion of \sgra, motions out of the plane are simpler to interpret.  
One needs only to subtract the small Z-component of the Solar Motion from the
observed motion of \sgrab to estimate the out-of-plane component of the peculiar 
motion of \sgra.  Using stars within about 0.1~kpc of the Sun and measured by 
Hipparcos, Dehnen \& Binney (1998) find the Z-component for the Solar Motion
to be $7.17 \pm 0.38~\kms$.
Other determinations of the Z-component of the Solar Motion, \eg 
$7.61 \pm 0.64~\kms$ for stars with distances out to a few kpc 
by Feast \& Whitelock (1997), are generally in agreement with the
result of \citet{DB98}, but have greater uncertainties.  Thus, we adopt
the \citet{DB98} value in this paper.
The motion of \sgrab in Galactic latitude, shown in Fig.~4, is re-plotted 
with an expanded scale in Fig.~5.  The dashed line in the figure is a weighted 
least-squares fit to the data.  The fitted slope of $-0.202\pm0.019$~mas~\peryr,
or $-7.6\pm0.7$~\kms for $\rnot=8.0$~kpc, agrees almost exactly with the reflex
of the Solar Motion out of the Galactic plane.   Allowing for a 0.1~degree
uncertainty in the tilt of the Galactic pole \citep{Blaauw60} increases slightly
the uncertainty in the out of plane motion from $0.7$ to $0.8$~\kms.
Subtracting $-7.17\pm0.38$~\kms from our measured apparent motion of 
\sgrab out of the plane of the Galaxy, we estimate the peculiar motion of 
\sgrab to be $-0.4\pm0.9$~\kms\ toward the north Galactic pole (see Table 3).

\begin{figure}
\plotone{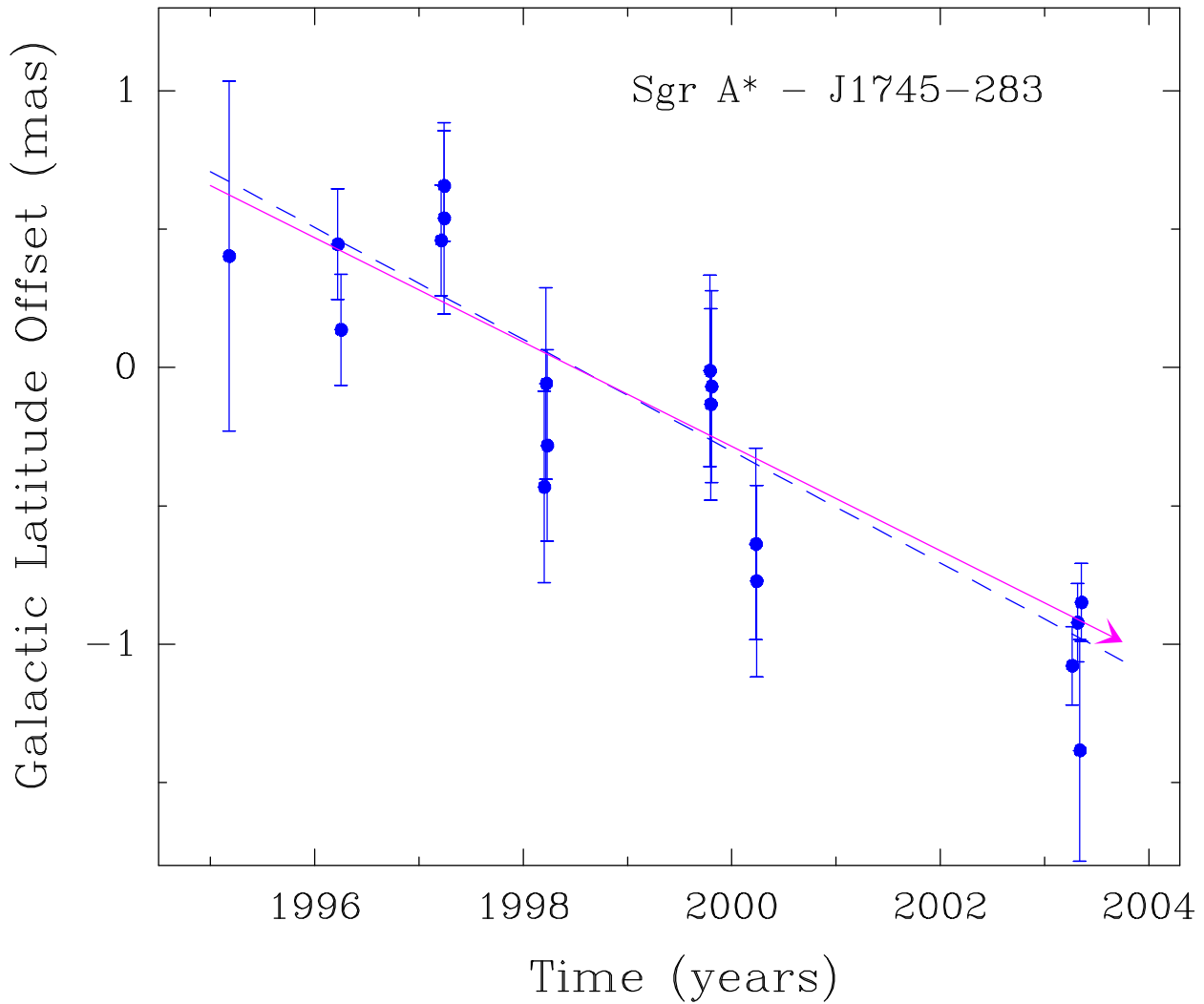}
\caption{\small 
Galactic latitude position residuals of \sgrab\ relative to 
J1745--283 from Fig.~4 on a finer scale.
The dashed line is the variance-weighted
best-fit proper motion of $-0.202\pm0.019$~mas~\peryr, or $-7.6\pm0.7$~\kms
for $\rnot=8.0$~kpc.  The solid arrow indicates
the apparent motion of \sgrab expected for the 7.17~\kms motion of
the Sun perpendicular to the plane of the Galaxy.
The slight tendency for data within a 1--2 week observing group
to correlate suggests a small source of systemic error ($\lax0.2$~mas)
for data prior to 2003, probably from unmodeled atmospheric delays.
Error bars are $\pm1\sigma$ and include an estimate of this systematic
uncertainty.   \label{fig5}}
\end{figure}

\subsection  {Acceleration of \sgra}

	The expected acceleration of the Sun in its orbit
about the Galactic center currently is undetectably small: 
$\tnot^2/\rnot\approx(220~\kms)^2/8.0~{\rm kpc}
   \approx6\times10^{-6}$~km~s$^{-1}$~y$^{-1}~,$
or in angular units $\tnot^2/\rnot^2\sim10^{-7}$~mas~y$^{-2}.$
Thus, unlike velocity measurements, which require precise knowledge 
and subtraction of the Sun's orbital contribution, 
any measurement of acceleration can be directly attributed to \sgra.  
Since the motion of \sgrab on the sky appears rectilinear, our
data can be used to set an upper limit on any apparent acceleration
of \sgra.  

Fitting the positions listed in Table~1 to a second-order
polynomial (\ie $x = x_0 + v_x \Delta t + 0.5 a_x \Delta t^2$)
allows an estimate of the acceleration.  We do not
detect any significant acceleration.  Acceleration uncertainties 
($1\sigma$) in equatorial components are $\sigma_\alpha=0.024,$
$\sigma_\delta=0.035$~mas~y$^{-2}$
and in Galactic components are $\sigma_l=0.035,$
$\sigma_b=0.026$~mas~y$^{-2}.$   
A value of 0.03 ~mas~y$^{-2}$ corresponds to $\approx 1$~km~s$^{-1}$~y$^{-1}$
at the distance of the Galactic center.
\citet{GR98} discuss the implications of upper limits to the acceleration
of \sgrab for the nature of \sgrab and Galactic structure constants.

\section {Limits on the Mass of \sgra}

	The orbits of stars as close as 100~AU from \sgrab require 
a central mass of about $4\times10^6$~\msun.  
Our limits on the proper motion of \sgrab can be used to indicate
how much of this mass must be directly associated with \sgra.  
Since the motion of \sgrab in the plane of the Galaxy has a much greater 
uncertainty (10--15~\kms, owing primarily to 
uncertainties in \rnot and \tnot) than the motion out of the plane of
the Galaxy (0.9~\kms), we focus mainly on the out-of-plane component.
We use our strong upper limit for \sgra's peculiar motion 
out of the plane of the Galaxy, in conjunction with 
1) the constraint that $\approx4\times10^6$~\msun\ of dark matter is 
contained in the central 100~AU region \citep{S03,Ghez03}, 
2) the observation that \sgrab is contained in this region 
\citep{MREG97,R03}, and 
3) the existence of $\sim10^6~{\rm to}~10^7$  stars orbiting the
galactic center within the central few parsecs \citep{Genzel03} 
to place firm lower limits 
to the mass directly associated with \sgra.

In the next sub-sections, we consider the possibility that \sgra's mass 
does not necessarily
dominate the region and compare the expected positions, velocity and
acceleration of \sgrab that might result with our observational limits.
We will first consider various possibilities for 
configurations of the $\approx4\times10^6$~\msun of material known to
occupy the central 100~AU region.  
The discussion divides along observational lines by a possible offset
of \sgrab from the dynamical center of a hypothetical dark matter 
distribution.  Since, \sgrab is observed to be within
100~AU of the center of this region \citep{S03,Ghez03,R03}, 
we do not consider offsets $>100$~AU.
In \S4.1, we consider offsets of \sgrab from the center of mass of the 
system between $4 < r < 100$~AU, where we would easily
detect changes in position, but not necessarily be able to
measure a velocity or acceleration, because the orbital period 
would be short compared to the time span of our observations
used to get accurate motions.  
In \S4.2, we consider offsets of $<4$~AU, where we could detect 
neither a positional offset nor \sgra's velocity or acceleration
even though the latter two quantities could be quite large.  

Following the discussion of the central 100 AU region, we 
calculate the effects of the $\sim10^6~{\rm to}~10^7$ stars,
believed to populate the inner two parsecs of the Galactic center region, 
on the motion of \sgra.  These stars orbit within the gravitational sphere
of influence of \sgrab (or more generally the $\approx4\times10^6$~\msun
of material contained within 100 AU of the center).  
Even for statistically isotropic distributions of
stars, random fluctuations in the mass distribution can result in a
significant motion for \sgra.
We show in \S4.3 that the gravitational pull of stars in the central parsecs
would induce a measurable motion for \sgra, were it not
to contain a significant fraction of the total mass in the region.
Similarly, we show in \S4.4 that a compact cluster of dark stellar remnants 
could also induce a significant motion for \sgra.  Finally, in \S4.5
we estimate a lower limit for the effects of stars beyond 2 pc from
\sgrab on its motion.

\subsection {\sgrab between $4 < r < 100$~AU of the Center}

Were \sgrab offset by more than 4~AU from the center of mass of the material 
in the inner 100~AU, we would expect to see detectable positional changes.
In Fig.~6 we plot the circular orbital speeds and periods of \sgrab for two 
dark matter masses: 1) where dark matter comprises essentially all of the 
$\approx4\times10^6$~\msun in the region and 2) where dark matter comprises
only 10\% of the mass in the region (requiring \sgrab to contain 90\%
of the mass).   For each case, we consider radial distributions of the
dark matter that follow power-laws, and we plot three distributions of 
density: falling as $1/r^2$, constant in $r$, and rising as $r^2$.
It is clear from the velocity plots that, for total dark matter 
mass of $>0.4\times10^6$~\msun and physically reasonable density
distributions, \sgrab would be orbiting at very high speeds.  
A similar conclusion holds for \sgra's acceleration.

However, the orbital periods (plotted in the lower panel of Fig.~6) 
would be less than 16~y for almost all models considered.  Since we measure 
the velocity (and acceleration) of \sgrab with data currently 
spanning 8~yr, we might not be sensitive to the very high velocities shown.  
For these cases, \sgra's orbit would result in strongly-aliased, quasi-random, 
positional shifts with an amplitude comparable
to the semi-major axis of its orbit.  The residual position offsets
of \sgrab (after removing the best-fit rectilinear motion) shown in Fig.~3
are well accounted for by measurement uncertainties, which are dominated
by unmodeled atmospheric propagation delays (see \S2 and a more detailed
discussion in Paper I).  The horizontal dashed lines in Fig.~3 indicate
a very conservative limit for positional ``noise'' from \sgrab orbiting the 
center of mass in this region.   Thus, components of \sgra's position cannot 
vary by more than 2 and 4~AU in the easterly and northerly 
directions, and we rule out \sgrab having an apocentric distance greater 
than these values.  Since we have position measurements in 2-dimensions, 
\sgra's orbital excursions are very unlikely to be hidden by projection effects.

\begin{figure}
\epsscale{0.65}
\plotone{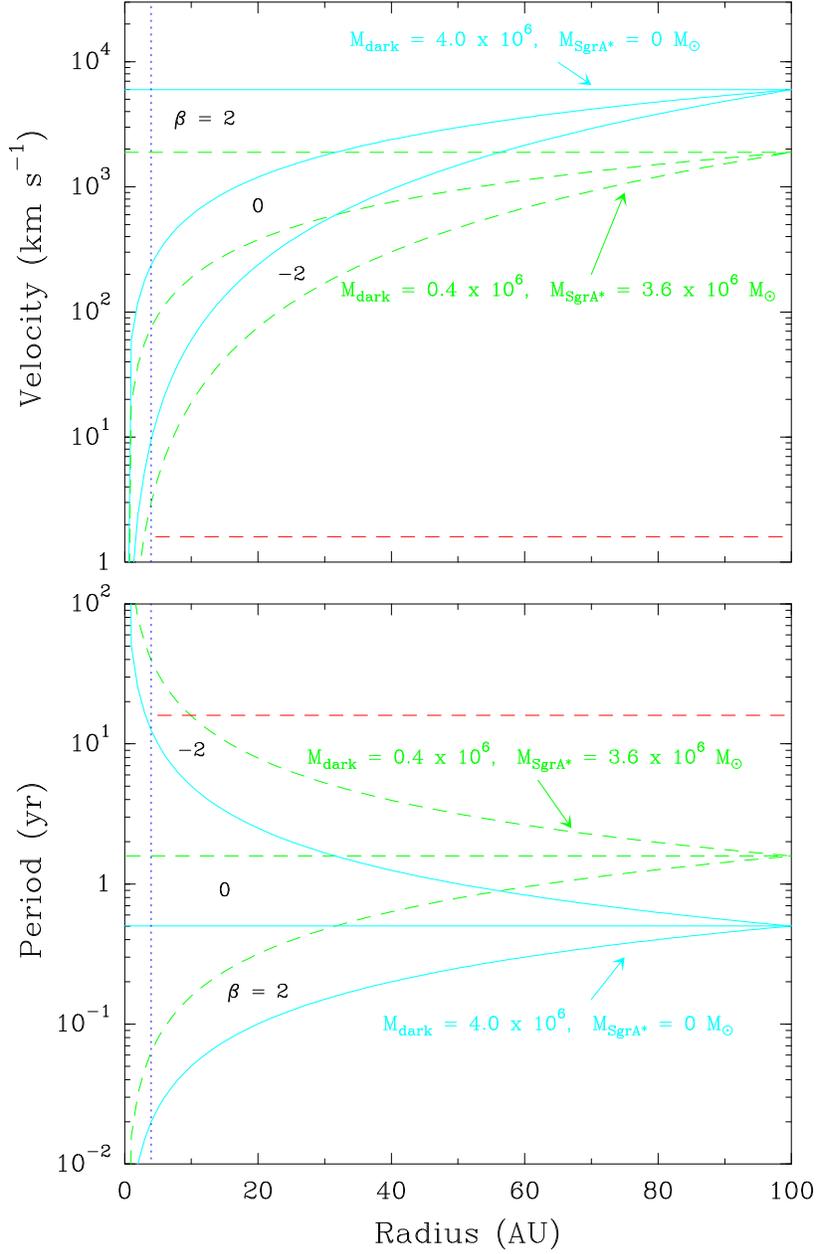}
\caption{\small 
Expected velocity ({\it upper panel}) and orbital period 
({\it lower panel}) of \sgrab owing to the gravitational force of 
hypothetical ``dark'' matter distributions as a function of radius.
A family of curves is plotted for each dark matter/\sgrab mass pair: 
{\it solid lines} are dark matter dominated and {\it dashed lines}
are \sgrab dominated.  The total mass is set to $4\times10^6$~\msun in 
central 100~AU region, as required by observed infrared stellar orbits.
For each dark matter mass, power-law distributions of density, $\rho$, 
with radius, $r$, given by $\rho \propto r^{-\beta}$ are plotted for 
$\beta = 2, 0, {\rm \&} -2.$  Horizontal dashed lines indicate our 
$2\sigma$ measurement uncertainty of 1.8~\kms for velocity and 16~y 
for the orbital period of \sgra, respectively;  
all values above the velocity limit would be ruled out, provided
the orbital period would be longer than the period limit.
Note that even though velocities of \sgrab can be very high,
orbital periods are very short and might not be detected (see \S4.1).
Vertical dotted lines indicate our lower limit of 4~AU for detection;
for radii less than this limit greater velocities are 
possible.  For radii greater than 4~AU position excursions for \sgrab
would be easily detected for any orbital periods.
 \label{fig6}}
\end{figure}

\subsection {\sgrab within $4$~AU of the Center}

If \sgrab remains within $4$~AU of the dynamical center of the
Galaxy, we might not detect any positional change, motion, or acceleration.
Indeed this is what one would expect if \sgrab is a SMBH containing 
most of the $\approx4\times10^6$~\msun in the central 100~AU.   
Could \sgrab be constrained to such a small region?
We will see in \S4.3 that the effects of stars in the central parsecs 
require \sgra, or whatever binds \sgrab to the center, most likely to have
greater than $\sim10^6$~\msun. 
Since the binding mass must be within 4~AU of center,
this requires an extraordinarily high mass density of 
$3\times10^{19}$~\msun pc$^{-3}$, 
almost surely requiring a SMBH (see \S5).  Thus, by assuming that \sgrab is
not a SMBH, we cannot avoid the conclusion that a SMBH occupies the Galactic
center.  

Further, postulating an extremely tight binary black hole is untenable as
it would would decay owing to gravitational radiation on timescales of 
$\sim (a_{\rm AU}^4/256f)$~y,
where $a_{\rm AU}$ is the separation of the black holes in AUs and $f$ is 
the ratio of the mass of the secondary to the total mass
in the binary \citep{ST83}.  Were \sgrab only a $10$~\msun object orbiting
a $4\times10^6$~\msun SMBH ($f=2.5\times10^{-6}$),
the time for gravitational decay from a $4$~AU radius is only $\sim4\times10^5$ y.
Thus, \sgrab would soon become (part of) a SMBH.  Of course, the far more likely
conclusion is that \sgrab is a SMBH.

\subsection    {Effects of Stars Within 2 pc of \sgra}

    While we have shown that \sgrab occupies the central 4~AU and contains 
enough mass to require a SMBH, it is worthwhile to investigate other constraints 
on \sgra's motion.   We now consider the effects of the orbital motions of
the large number of stars observed in the central stellar cusp
on \sgra.  Random distributions of stars in orbit about \sgra\ will 
produce small asymmetries in the mass distribution, which vary over time
and lead to small motions of \sgra\ about the center of mass of the
system.  In the following we first develop a simple analytic 
estimate of the effects of stars of a single mass in circular
orbits on the motion of \sgra.  After this we present results of
detailed numerical simulations of the effects of $\sim10^6~{\rm to}~10^7$ stars
in the central stellar cusp within $2$~pc of \sgra.
The numerical simulations allow us to investigate the effects of
a distribution of stellar masses and orbital eccentricities.

\subsubsection {Analytic Approach}

    We now consider the effects of bound orbital motions of
a large number of stars on the motion of \sgra.  
Assume an isolated system with \sgrab at the center of mass of a spherically 
symmetric random distribution of orbiting stars.  
Without loss of generality, we place the center of mass at the origin of 
a coordinate system.  For this closed
system, the center of mass remains fixed and hence
$$M {\vec V} = -\sum_i m_i {\vec v_i}~~,\eqno(1)$$
where $M$ and $\vec V$ are the mass and velocity of \sgra, and 
$m_i$ and $\vec v_i$ are the mass and velocity of the $i^{\rm th}$ star.
Squaring Eq.~(1) yields
$$M^2 V^2 = \sum_i {m_i^2 v_i^2} + \sum_i\sum_{j\ne i} {m_i m_j 
   {\vec v_i}\bullet{\vec v_j} } ~~.\eqno(2)$$
Taking expectation values for random distributions of stellar velocities,
the cross terms vanish and we find
$$M^2 <V^2> = \sum_i {<m_i^2 v_i^2>}~~.\eqno(3)$$
For simplicity, assume a single characteristic stellar mass, $m$,  and
replace the summation  in Eq.~(3) with an integral:
$$\sum_i {<m_i^2 v_i^2>} = m^2\int_r v^2(r) \Bigl({dN\over dr}\Bigr) dr~~,$$
where $\bigl({dN\over dr}\bigr) dr$ is number of stars between radii of 
$r$ and $r + dr.$  Eq.~(3) then becomes 
$$<V^2> = {m^2 \over M^2} \int_r v^2(r) \Bigl({dN\over dr}\Bigr) dr~~.\eqno(4)$$

For a power-law distribution of stellar mass density with radius
given by $\rho(r) = \rho_0 (r/r_0)^{-\alpha},$
$$\Bigl({dN\over dr}\Bigr) dr= (\rho/m)4\pi r^2 dr =
N_0 \Bigl({r \over r_0}\Bigr)^{2-\alpha}~d(r/r_0)~~,\eqno(5)$$
where $N_0 = (\rho_0/m)4\pi r_0^3.$
For circular stellar orbits about a fixed central potential (at the
average position of \sgra),
$$v^2(r) = {G M_r \over r}~~,\eqno(6)$$
where $M_r$ is the total mass enclosed within $r$, given by
$$M_r = M + \int_0^r \rho(r) 4\pi r^2 dr~~.\eqno(7)$$
Defining $\xi = r/r_0$ and $M_0 = \rho_0 4\pi r_0^3 / (3-\alpha)$
and inserting Eqs.~(5-7) into Eq.~(4) yields
$$<V^2> = {Gm^2 N_0 \over M r_0} \int_\xi 
   \Bigl( 1 + {M_0\over M} \xi^{3-\alpha} \Bigr) \xi^{1-\alpha} d\xi~~.\eqno(8)$$
Integrating Eq.~(8) results in
$$<V^2> = {Gm^2 N_0 \over M r_0} 
   \Bigl[ {\xi^{2-\alpha}\over{2-\alpha}} 
        + {M_0\over M} {\xi^{5-2\alpha}\over{5-2\alpha}} 
   \Bigr]_{\xi(r_{min})}^{\xi(r_{max})}
    ~~,\eqno(9)$$
for $\alpha\ne2~{\rm or}~2.5.$ 
(For $\alpha=2~{\rm or}~2.5$ replace the appropriate ``$\xi^x/x$'' terms with
$\ln{\xi}$.)

Eq.~(9) has the interesting property that it transitions smoothly from
equipartition of momentum to equipartition of kinetic energy as the total
mass in stars goes from that of a single star to a value equal to that of \sgra.
This can be seen by noting that Eq.~(9) can be approximated by
$$<V^2> ~\sim~ {Gm^2 N_0 \over M r_0}~~,\eqno(10)$$ 
multiplied by a dimensionless quantity of order unity.
For a single star orbiting \sgra, $m N_0 = m$ and Eq.~(10) becomes
$$M^2 <V^2> ~\sim~ m^2 {G M \over r_0}.$$ 
Since the quantity $G M / r_0$ equals the expected square of the speed, $<v^2>$, 
of a star orbiting \sgrab at a radius $r_0$, we find equipartition of momentum: 
$$M <V^2>^{1/2} ~\sim~ m <v^2>^{1/2}.$$ 
However, when the total stellar mass equals that of \sgra, $m N_0 = M$
and Eq.~(10) yields equipartition of kinetic energy:
$$M <V^2> ~\sim~ m <v^2>.$$  Thus, the combined effects of bound
stellar orbits result in equipartition of kinetic energy, when the total mass
in stars equals the mass of the dominant central object.
This resolves the problem, raised in Paper I, regarding the applicability of 
equipartition of kinetic energy for stellar systems bound to a massive object.

We evaluate Eq.~(9) for the broken power law distribution
of stars in the central cusp observed by \citet{Genzel03}:
$\rho(r) = 1.2 \times 10^6 (r/0.4~{\rm pc})^{-\alpha}~~\msun~{\rm pc}^{-3},$
where $\alpha=1.4$ for $r<0.4$~pc and $\alpha=2.0$ for $r\ge0.4$~pc.
This stellar mass distribution contains $4\times10^6$~\msun within
$r\approx2$~pc of \sgra.  We do not continue beyond
this radius, as the stellar mass would start to dominate and our assumption
that stars are bound to \sgrab starts to break down.
Assuming a characteristic stellar mass of $m=0.453$~\msun,
equal to the first moment of a standard initial mass function \citep{Allen00},
and \sgrab contains $4\times10^6$~\msun, we find ${<V^2>}^{1/2}=0.07$~\kms,
which implies individual component speeds of 0.05~\kms.

The mass function of stars in the inner two parsecs is likely to be flatter
than a standard IMF, owing to mass segregation effects.  
Since the expected motion of \sgrab scales 
as $\sqrt{m}$, we would expect a higher rms speed for \sgrab than calculated
above, possibly by a factor of two or more.   
Overall, our estimate of the motion of \sgrab is {\it extremely} conservative.
The assumptions of 1) a standard IMF,
2) a perfectly random stellar distribution (no clumping or anisotropies), 
3) no contribution from a possible cluster of dark stellar remnants
close to \sgrab (see \S4.4), and 4) no 
contribution from mass asymmetries beyond $r=2$~pc (see \S4.5),
all contribute to give the lowest possible motion for \sgra.

\subsubsection {Direct Simulations}

The analytical approach of the previous section assumes 
stars of a single mass and circular orbits.  
In reality, one expects a distribution of stellar masses and, 
especially in the crowded environment of such a 
dense stellar system, a wide distribution of orbital eccentricities.
In order to understand better the effects of stars in the central cusp
on the motion of \sgra, we carried out full numerical 
simulations of the effects of the $\sim10^6~{\rm to}~10^7$ stars 
thought to orbit within 2~pc of the Galactic center.  

We are interested in the change in position of
\sgrab over a time period of eight years.  Over
such a short time span (compared to typical stellar orbital periods
of $\sim10^3$~y at 0.1~pc radii), stellar motions are very short orbital 
arcs.  Thus, there is no need to include the gravitational effects
of individual stars on each other as is done in N-body calculations; 
we can assume that stars move along orbits determined by the mass enclosed 
within their semi-major axes.  By avoiding full N-body calculations, 
we are ignoring any collective effects from stellar clumping,
which would increase the expected speed of \sgra. 
So, we are being very conservative when 
applying the results to obtain a lower limit to the mass of \sgra.

We initialized a simulation by assigning each star a random stellar 
mass, consistent with a chosen stellar mass function.  We evaluated 
three stellar mass functions: a standard IMF \citep{Allen00}, 
one flatter by 0.5 in the power-law indexes, and a second flatter by 
1.0 in the power-law indexes.
Next, we assigned each star a random semi-major axis, distributed
as a broken power-law in radius as described by \citet{Genzel03}
and listed in the previous section.
Orbital eccentricities were chosen from a uniform random
distribution with values between 0 to 0.999.
Tests with different distributions of eccentricities (\eg $dN/de \propto e$
and $dN/de \propto \sin{e}$ produced nearly the same results.

Each semi-major axis was initially
placed on one Cartesian axis and the orbits randomized in space
by three Euler rotations with angles chosen from uniform
random distributions.  
Note that we do not expect a significant departure from spherical symmetry
within 2 pc of the Galactic center.
Finally, we assigned each star a uniformly distributed
initial mean anomaly. 

Stars were assumed to orbit about a dominant central mass.  
The position of each star at time, $t$, 
was determined from its orbital elements by solving Kepler's equations. 
The center of mass of the stellar cluster at time zero
was determined and \sgrab placed at that position.  The position of 
\sgrab at any other time could then be computed directly from the 
stellar masses and positions by requiring that the center of mass of 
the entire system remain fixed.  The full velocity vector for
\sgrab was calculated by differencing positions for times separated
by 8~y.  In test simulations, the change
of position of \sgrab was linear, with no detectable jitter,
over this time span.  Over longer periods of time of order $10^4$~y,
\sgrab slowly wandered by $\sim100$~AU, responding to changes in the
center of mass of the surrounding stars at characteristic distances 
of 1~pc.  (This is unlikely to affect measurements of stellar orbits
for stars within $\approx0.01$~pc, as these stars are tightly bound
to \sgrab and should wander with it.)

\begin{figure}
\epsscale{0.80}
\plotone{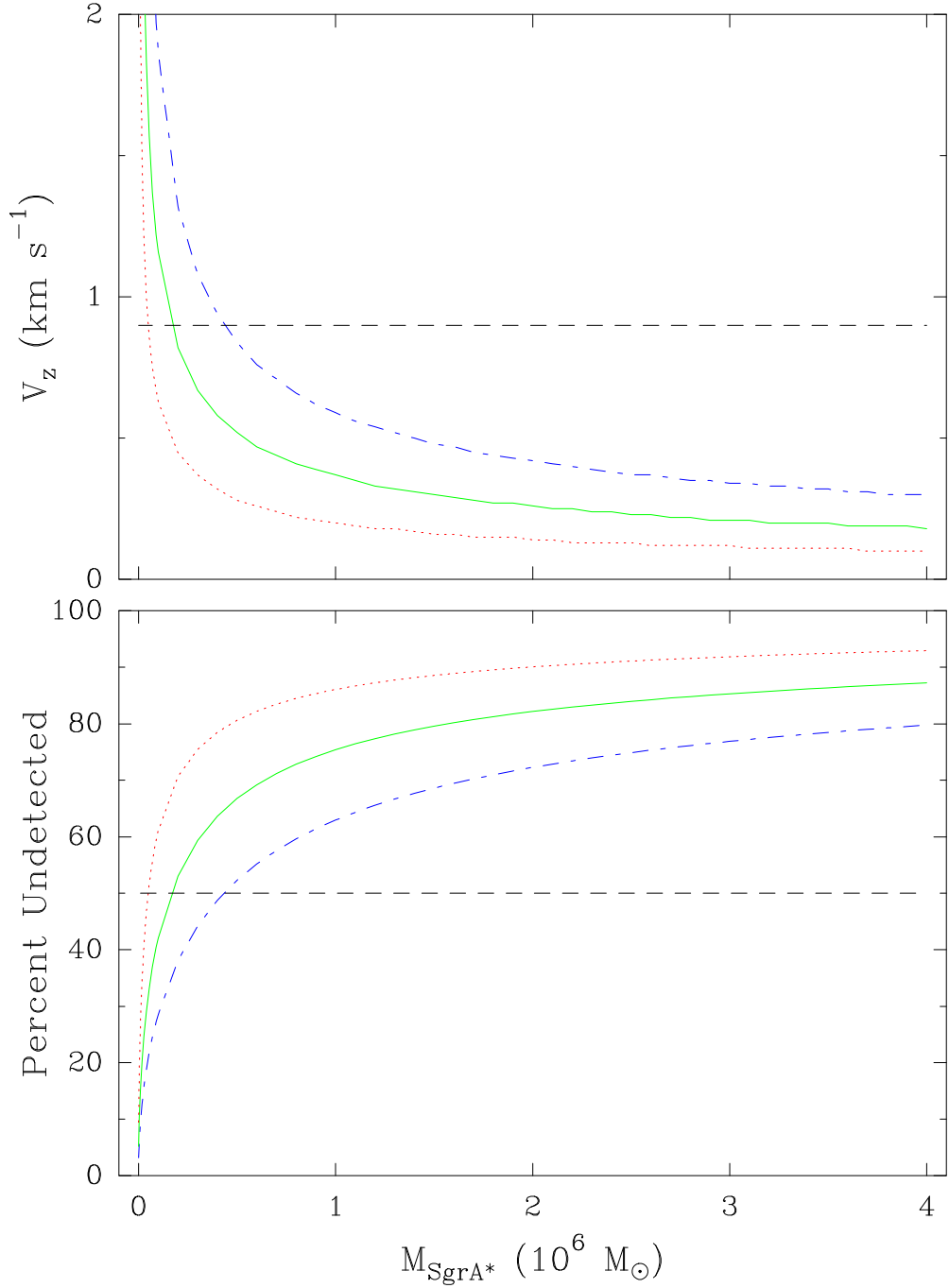}
\caption{\small
Results of direct simulations of the motion of \sgra,
owing to the gravitational forces of stars in random orbits 
within 2~pc of the Galactic center.
Three simulations for different stellar initial mass functions (IMF)
are plotted:
the {\it dotted} line corresponds to a standard IMF, and the
{\it solid,} and {\it dot-dashed} lines correspond to  
mass functions flatter by 0.5 and 1.0 in their power-law indexes.
A flatter than standard IMF is expected owing to mass segregation from
dynamical friction and observed in the Arches Galactic center cluster.
{\it Upper panel:} The simulated root-mean-square 
of one component of the velocity of \sgrab as a function of its mass.  
The {\it dashed line} indicates our $1\sigma$ measurement uncertainty of 0.9~\kms.
{\it Lower panel:} The percent of simulations
where the motion of \sgrab is less than a trial measurement, with the
{\it dashed line} indicating the 50\% or maximum likelihood limit.
 \label{fig7}}
\end{figure}

We computed 1000 simulations of the effects of $\sim10^6~{\rm to}~10^7$ stars
on the motion of \sgrab for the three stellar mass functions (discussed above). 
For a standard initial mass function (IMF) and a mass of 
\sgrab of $4\times10^6$~\msun, the root-mean-square speed for \sgrab
was 0.17~\kms and each component of the velocity vector 
was 0.10~\kms.  This single component speed is 
a factor of two greater than the 0.05~\kms value obtained
by the analytic treatment in \S4.3.1.  However,
as pointed out earlier, that treatment assumed circular orbits and
a single (average) stellar mass.  
We tested our numerical simulation under these restrictive conditions 
($e=0$ and $m=0.453$~\msun, corresponding to the first mass moment of a 
standard IMF) and obtained
0.05~\kms for a single component speed of \sgra, in agreement
with the analytical result.  

In order to determine a lower limit to the mass of \sgra, given
our 0.9~\kms ($1\sigma$) measurement uncertainty in the component of
\sgra's velocity out of the plane of the Galaxy, we estimate the
motion of \sgrab as a function of its mass.
As shown in \S4.3.1, equipartition of kinetic energy holds when
the total mass in stars equals that of \sgra.  Thus, the expected 
motion of \sgrab ($V_z$) should scale as $1/\sqrt{M}$, with one
possible caveat.
For cases where $M \ll 4\times10^6$~\msun, it would be important to
replace the ``missing mass'' in order to satisfy the central mass
constraints from infrared stellar orbits.  Not doing this might violate the
assumption in our calculations that \sgrab is perturbed only
by the observed central stellar cluster.  One could correct for this
with full N-body calculations, provided one knew the nature of this
hypothetical ``missing dark matter.''  However, this is not known.
Note that if dark matter 
(other than a SMBH) dominates in the central 100~AU region, then, 
as discussed in \S4.1, \sgrab must still be bound within 4~AU of center.

The motion of \sgrab as a function of its mass is shown in Fig.~7.  
The upper panel gives the rms speed for one component of \sgra's motion.
The three curves correspond to three different stellar mass functions.
While we do not know the stellar mass function of the Galactic center 
cluster, it is observed to contain an unusually large 
population of very massive stars \citep{K91,N97}.  This may result 
from the unusual physical conditions in the Galactic center, 
which might favor high mass star formation \citep{M93}.  
Also, the Galactic center cluster is very likely to have undergone significant mass
segregation, with more massive stars ``sinking'' closer to the
center than less massive stars, owing to the effects of dynamical
friction.  Observationally, the Arches cluster near the Galactic Center
has a mass spectrum index flatter by about +0.6 to +0.8 than other 
Galactic clusters \citep{F99,SGB02} or a standard IMF.  Because of
these findings, we feel the best estimate of \sgra's expected motion
is between the curves for stellar mass function whose indexes are flatter than
for a standard IMF by 0.5 to 1.0. 
Thus, if \sgrab contains $4\times10^6$~\msun, we expect it to have a 
motion in one-dimension of between 0.18 and 0.30~\kms, respectively.   
Decreasing the mass of \sgrab would increase its expected speed as shown in Fig.~7.

We now proceed to estimate a maximum-likelihood lower limit to the mass of
\sgra, by calculating the percent of stellar cluster simulations for which a velocity
component of \sgra's motion would be less than a measurement
drawn from a Gaussian random distribution with $\sigma_m=0.9$~\kms,
matching our observational accuracy.  For any given measurement of the 1--dimension  
velocity of \sgra, $V_m$, the fraction of the simulations that would
have a lower speed is given by 
$${\rm Erf}(V_m,\sigma_s) = {1\over \sigma_s \sqrt{2\pi}}~
    \int_{-V_m}^{-V_m} e^{-V^2/2\sigma_s^2} dV~~,$$
where $\sigma_s$ is the root-mean-square of the simulated (1-D) velocities of \sgra.  
For a Gaussian distribution of measurements, $V_m$, the probability of measuring a 
value between $V_m$ and $V_m + dV$ is 
given by, 
$$G(V_m,\sigma_m) dV_m = {1\over \sigma_m \sqrt{2\pi}}~e^{-V_m^2/2\sigma_m^2} dV_m~~.$$
Then the probability, $P(\sigma_s,\sigma_m)$, that we
would not detect a simulation of \sgra's motion with a given measurement
(\ie $|V_s| < |V_m|$) is given by
$$P(\sigma_s,\sigma_m) = \int_{-\infty}^\infty {\rm Erf}(V,\sigma_s)~G(V,\sigma_m)~dV~~.\eqno(11)$$

In the lower panel of Fig.~7 we plot $P(\sigma_s,\sigma_m)$ as a function of
the mass of \sgrab for the three stellar mass functions. 
We find that simulated
motions of \sgrab would exceed measurements 50\% of the time
for \sgrab masses of $0.05,~0.2, {\rm and}~0.5\times10^6$~\msun for a standard IMF,
and stellar mass function indexes flatter by 0.5 and 1.0, respectively.
These mass estimates are maximum-likelihood lower limits to the mass of \sgra.
Because the stellar mass function is likely considerably flatter than a standard IMF 
(see discussion above), we adopt a lower limit to the mass of 
\sgrab of $0.4\times10^6$~\msun, a value between those for the two flatter 
stellar mass functions.

As discussed in \S3, the component of the motion of \sgrab in the plane of
the Galaxy (in the direction of Galactic rotation) is much less well 
determined than the component out of the plane.  
Because we can effectively use only one component of the velocity vector to
limit the mass of \sgra, there is non-negligible possibility that projection
effects could hide a more significant motion.  An examination of Fig.~7 reveals
that, as one demands higher confidence for the percentage of cases for which we would
not detect a motion of \sgra, the mass limit decreases rapidly.
For example, for a 95\% confidence limit ($<5$\% undetected in Fig.~7), 
the mass limit is $\sim10^3$~\msun.  However, reducing the mass of \sgrab
well below $\sim10^6$~\msun opens the question of how to satisfy the
constraint from stellar orbits that $4\times10^6$~\msun is contained 
in the central 100~AU.  Replacing most of the mass in the central 100~AU 
with dark matter \citep{BMTV02} or stellar remnants may already be ruled 
out \citep{M98,Ghez03,S03}.
Because of these difficulties, as well as the extremely conservative 
assumptions used to calculate the motion of \sgra, we adopt the 
maximum likelihood estimate of $0.4\times10^6$~\msun for the lower limit to the
mass of \sgra.

\subsection     {Effects of a Central Dark Stellar Cluster}

In addition to a visible stellar cluster, the center of the Galaxy 
may contain large numbers of dark stellar remnants.
\citet{M04} suggest that the infrared positional
data for star S2 support a model with a SMBH of $3.7\times10^6$~\msun,
plus a compact dark cluster with a mass of  $0.4\times10^6$~\msun.
Such a compact dark cluster might be composed of massive stellar remnants
which migrated toward the Galactic center over billions of years owing
to dynamical friction.  If such a dark cluster exists, it would also
affect the motion of \sgra.  

In order to quantify the motion of \sgrab under the influence of
a compact cluster of stellar remnants, we modified the computer
code used in the previous section to simulate this case.  Instead
of power-law radial distributions, we adopted Plummer distributions
following \citet{M04}.  We considered dark clusters
of stellar remnants totaling $0.4\times10^6$~\msun, containing 
50\% neutron stars of 1.4~\msun and 50\% black holes with a 
uniform distribution of mass between 3 and 25~\msun.  The
core radius of the cluster was varied and large numbers of 
random simulations were evaluated to determine the rms velocity 
of a $3.6\times10^6$~\msun \sgra.   As before, we calculated
the position of \sgrab at times separated by 8 years and
differenced these to determine a velocity in order to mimic our
observations.  

\begin{figure}
\epsscale{0.85}
\plotone{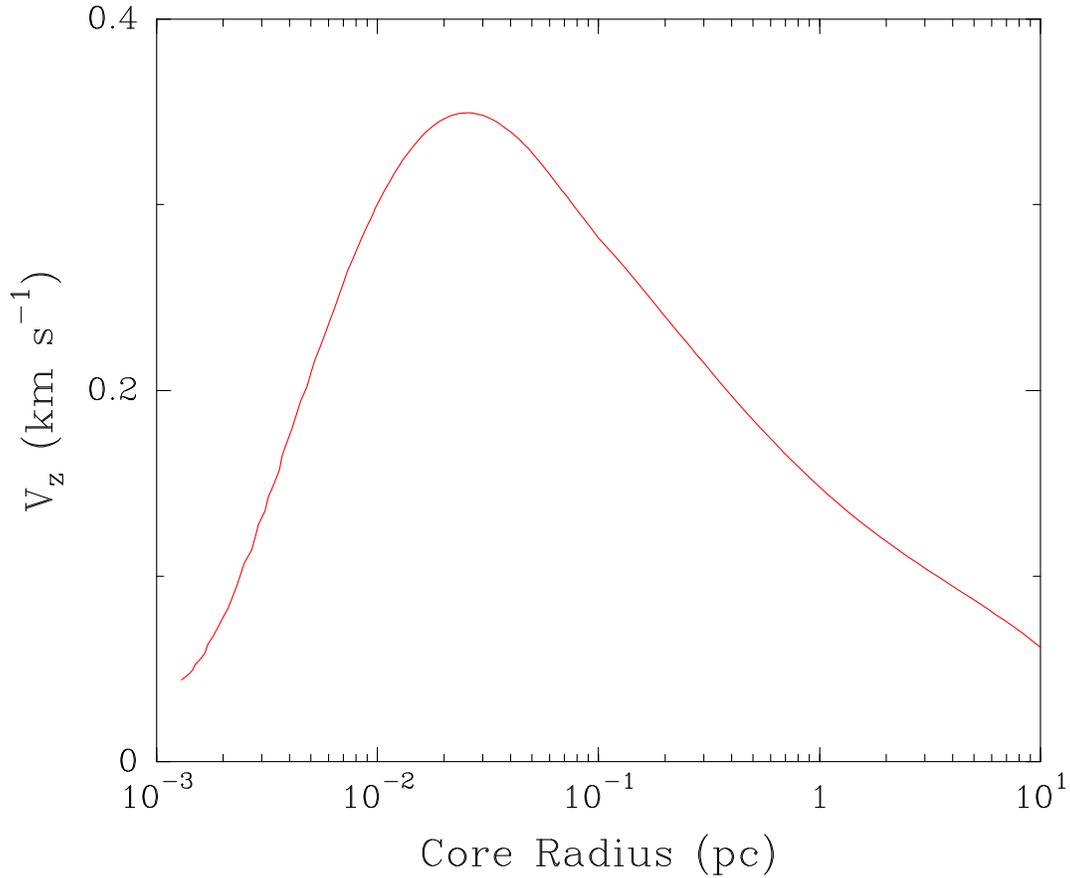}
\caption{\small 
Results of direct simulations of the motion of \sgra,
owing to the gravitational forces of random stellar orbits, for
a hypothetical cluster of stellar remnants.  The stellar
cluster was assumed to contain $0.4\times10^6$~\msun with
Plummer radial distribution.
The line traces the simulated root-mean-square 
of one component of the velocity of \sgrab as a function of 
the core radius of the Plummer distribution.   For core
radii less than about 0.03~pc, the velocity obtained by fitting
a straight line to positions over an 8 year period, similar to 
our observations, is smaller than the instantaneous velocity
owing rapid fluctuations in position caused by stellar remnants with orbital
periods $<16$~y.
 \label{fig8}}
\end{figure}

The results of our simulations are shown in Fig.~8.  A cluster of only  
$\sim5\times10^4$ dark stellar remnants containing $0.4\times10^6$~\msun
could contribute $>0.3$~\kms\ to one component of the measured motion of \sgra.
\citet{M04} find a best-fit core radius of 0.0155~pc, which
would give a $3.6\times10^6$~\msun \sgrab a 1-dimensional rms motion of 0.34~\kms.
Furthermore, over a wide range of core radii from about 0.004 to
0.4~pc, such a cluster could contribute about $>0.2$~\kms.  Thus, a
contribution to the motion of \sgrab from a cluster of dark
stellar remnants might be comparable to the contribution of the
$\sim10^6~{\rm to}~10^7$ stars observed within 2~pc of \sgra.
Were we to add a contribution from such a cluster of dark stellar
remnants ($\approx0.3$~\kms) to the motion of \sgrab from the calculations in \S4.3
($\approx0.2-0.3$~\kms), the quadrature sum for the 1--dimensional
motion of \sgrab would be $\approx0.4-0.5$~\kms.
This would increase the lower limit of the mass of \sgrab to 
0.7 or $1.7\times10^6$~\msun\ for the two flatter mass functions considered above.

\begin{deluxetable}{lllllll}
\tablenum{4}
\tablewidth{0pt}
\tabletypesize{\small}
\tablecaption{Mass Densities of SMBH Candidates}
\tablehead{
\colhead{Source}         & 
\colhead{Density}    & \colhead{Mass} &
\colhead{Radius}    & \colhead{References} &
 \\
\colhead{}               & 
\colhead{(\msun~pc$^{-3}$)}       & \colhead{(\msun)}  &
\colhead{}       & \colhead{}  &
 }
\startdata
\\
M~87 ............  &$\pg1\times10^5$    &$\pg2\times10^9$    &\pl18~pc     &1 \\
NGC~4258 ...       &$\pg7\times10^9$    &$\pg3\times10^7$    &\pl0.1~pc    &2 \\
\sgrab .........   &$>8\times10^{15}$   &$\pg4\times10^6$    &$<100$~AU    &3 \\
\sgrab .........   &$>7\times10^{21}$   &$>4\times10^5$      &$<0.5$~AU    &4 \\
SMBH .........     &$\pg2\times10^{25}$ &$\pg4\times10^6$    &\pl0.08~AU   & \\
\enddata

\tablenotetext{1}{\citet{F94,H94}}
\tablenotetext{2}{\citet{M95}}
\tablenotetext{3}{\citet{S03,Ghez03}}
\tablenotetext{4}{Mass limit from this paper; Size limit from 
\citet{Rogers94,Krichbaum98,D01,B04}}

\end{deluxetable}

\subsection     {Effects of Stars Beyond 2 pc}

The large number of stars (as well as a significant amount of dust and gas)
at radial distances greater than 
2~pc from \sgrab must also contribute to the motion of \sgra.
However, since the total mass of such material can greatly exceed
that of \sgra, we cannot use the methods of \S4.3, which
require the stars to reside in the gravitational sphere of influence
of \sgra.  In general, this problem requires very large N-body
calculations.  
Recently, \citet{CHL02} and \citet{Dor03} addressed this theoretically
as a Brownian motion problem and by N-body simulations.
Both papers conclude that the ``Brownian particle,'' \sgra, would
come into equilibrium rapidly with the sea of surrounding stars 
and achieve equipartition of kinetic energy.  

For equipartition of kinetic energy, 
$$M <V^2> \approx m <v^2>~~,\eqno(12)$$
where capital and lower case symbols refer to \sgrab and 
a typical star, respectively.
The characteristic speed of a star is given by 
$<v^2> \approx G M_r / r,$ where $M_r$ is defined in Eq.~(7).
If the stellar density is given by $\rho = \rho_0 (r/r_0)^{-2},$
then $M_r = M + \rho_0 4 \pi r_0^2 r.$
At a sufficiently large radius such that stars dominate
the enclosed mass, $M_r/r$ approaches a constant and
$<v^2> \approx G \rho_0 4 \pi r_0^2.$
For this case, Eq.~(12) yields
$$<V^2> \approx {m \over M} G \rho_0 4 \pi r_0^2~~,\eqno(13)$$
independent of radius.
As before, adopting the \citet{Genzel03} model for the radial distribution
of stars with $r>0.4$~pc,
$\rho(r) = 1.2 \times 10^6 (r/0.4~{\rm pc})^{-2}~~\msun~{\rm pc}^{-3},$
Eq.~(13) indicates that distant stars with $m\approx1$~\msun
should contribute to the motion of a $4\times10^6$~\msun black hole
as $<V^2>^{1/2} \approx 0.05$~\kms.   This speed should be considered
a lower limit, since any asymmetries caused, for example, by clustering of stars
would increase the expected motion of \sgra, as would inclusion of a expected
stellar mass function.  Also,
as pointed out by \citet{BS99}, the possibility of large scale
asymmetries in not only the stellar distribution, but also 
from massive molecular clouds, could significantly increase
the motion of \sgra.  Note, however, that very large scale asymmetries 
(\eg $>100$~pc from the center) would most likely be
confined to the plane of the Galaxy and may not affect the out-of-plane
motion of \sgra.

\section        {Is \sgrab a Super-Massive Black Hole}

Based on the results discussed in \S4,  
we conclude that the compact radio source, \sgra, contains
at least 10\% of the $4\times10^6$~\msun in the inner 100~AU region.   
The apparent size of \sgrab
measured with VLBI techniques is dominated by scatter broadening, 
and its intrinsic size is less than about 1~AU \citep{Rogers94,Krichbaum98,D01,B04}.
Thus, we require a mass of $>0.4\times10^6$~\msun within a radius of $<0.5$~AU,
which yields an astounding mass density of $>7\times10^{21}$~\msun~~pc$^{-3}.$

In making the density calculation, we have assumed that the size of
the emitting region is equal to or greater than the size of the
region containing the mass.  This is true for almost all astrophysical
sources.  Notable exceptions for radio sources are solar flares
and pulsars.  However these sources are sporadic (either
flares or pulses) and are usually characterized by gyro-synchrotron
emission with very step spectral indexes.  \sgrab does not share
these characteristics, especially in the radio band where its
emission is dominated by a generally slowly varying, smooth, rising spectrum.
Thus, we conclude that our assumption that the
radiative size of \sgrab equals or exceeds the physical size is reasonable.

Table 4 lists the mass densities of notable super-massive black hole candidates.
The mass density we have determined for \sgrab is six orders of
magnitude higher than can be determined currently from infrared stellar orbits.
It is 12 orders of magnitude greater than for NGC~4258
(from imaging an 0.1~pc radius accretion disk through its H$_2$O maser emission)
and nearly 17 orders of magnitude greater than for M~87 (from HST spectroscopy
of a central stellar cluster).
Our mass density is within about three orders of magnitude of that for a 
$4\times10^6$~\msun black hole within its Schwarzschild radius, 
$R_{\rm Sch}$.  
If one adopts a minimum possible radius of $3R_{\rm Sch}$ for a stable
object, then the lower limit for the mass density of \sgrab is
within only two orders of magnitude of the super-massive black hole value. 
Thus, the mass density determined from the proper motion limit for \sgra,
coupled with its very small intrinsic size,
is the strongest and most direct evidence to date that the compact 
radio source, \sgra, is a super-massive black hole.

\section{Limits on a Binary Black Hole}

The possibility that a binary black hole exists in the center of the 
Galaxy has been discussed by \citet{HM03} and \citet{YT03}.  
Such a binary could arise as a dense stellar cluster, containing an 
intermediate mass black hole, is dragged toward a more massive black hole 
(\sgra) at the center of the Galaxy.   This scenario is interesting as it
might help to explain the existence of large numbers of massive young stars 
in the central stellar cluster, which are unlikely to have formed
at their present locations.   

As suggested by \citet{HM03}, limits on the orbital excursion and 
velocity of \sgrab provide strong constraints on the masses
and semi-major axes of secondary (intermediate mass) black holes
in the Galactic center region.  Our current observations support limits 
between those plotted by \citet{HM03} in their Fig.~2 for 
``astrometric resolutions'' of 0.1 and 1.0~mas.  
The limits on these parameters are complex and depend on both
parameters, but roughly we can 
exclude secondary black holes with masses greater than $\sim10^4$~\msun\ 
and semi-major axes between $10^3$ and $10^5$~AU from \sgra.  Excluding stellar
mass black holes ($<100$~\msun) will require more than an order
of magnitude better astrometric accuracy and is unlikely in the near
future.  

\section	{Conclusions}

We have measured the position of the compact non-thermal 
radio source, \sgra, at the center of the Galaxy relative to
extragalactic radio sources.  The apparent motion of \sgrab is consistent
with that expected from the orbit of the Sun around the Galactic center.
Any peculiar motion of \sgrab perpendicular to the plane of the Galaxy
is less than 1.8~\kms ($2\sigma$).  This result is complementary to
infrared observations of stellar orbits 
at the Galactic center, which require $4\times10^6$~\msun within a radius
of 100~AU of \sgra.  
The results of several different analyses indicate a significant fraction,
if not all, of the mass in the central 100~AU is tied to \sgra. 
Were \sgrab not a SMBH, it must be bound to the inner 4~AU of the 
dynamical center of the Galaxy.  This would imply an extraordinarily 
high mass density and probably require a SMBH.  

The gravitational attractions of the $\sim10^6~{\rm to}~10^7$~stars 
within 2~pc of the Galactic center impart a significant motion to \sgra.  
Based on numerical simulations of the central star cluster and our upper limit 
to the motion of \sgrab out of the plane of the Galaxy, 
a maximum-likelihood lower limit for the mass of \sgrab is $0.4\times10^6$~\msun.  
These analyses make very conservative assumptions that would
tend to underestimate the motion of \sgrab and, hence, underestimate
the mass limit.  This is the first {\it direct} evidence that
a compact radiative source at the center of a galaxy is a super-massive
object.  Other measurements determine a large mass, but can only {\it indirectly} 
associate it with the radiative source through positional agreement.  

The observed radio frequency size of \sgrab is less than 1~AU,
after accounting for the effects of interstellar scattering.  
The mass density implied by having at least $0.4\times10^6$~\msun
within a 0.5~AU radius is a staggering $7\times10^{21}$~\msun~pc$^{-3}$!
This is only about 3 orders-of-magnitude lower than the mass
density of a $4\times10^6$~\msun black hole within its Schwarzschild
radius, providing overwhelming evidence that \sgrab is a super-massive
black hole.
Should future VLBI measurements at 1~mm wavelength show that
the intrinsic size of \sgra\ is $\approx0.2$~AU, then we could 
conclude that most of the mass in the region required for 
a super-massive black hole is contained within a few $R_{Sch}$ for \sgra.

\acknowledgments
We thank V. Dhawan for helping with the VLBA setup, A. Loeb and
F. Rasio for discussions that led to the analysis presented in
\S4.3.1, and J. Goodman and R. Narayan for discussions on
the characteristics of the probability distribution for the mass limit for \sgrab 
using only one component of its motion.

\appendix
\section {Appendix}
The IAU Galactic plane \citep{Blaauw60} 
is defined in B1950 coordinates by a north Galactic pole toward 
$\alpha_p^\fifty\equiv12^{\rm h}~49^{\rm m}, \delta_p^\fifty\equiv+27.4^\circ$
and the ``zero of longitude is the great
semi-circle originating at the new north galactic pole at the position 
angle $\theta^\fifty\equiv123^\circ$.''   This gives a B1950 origin for galactic
coordinates of $\alpha_0^\fifty=17^{\rm h}~42^{\rm m}~26.603^{\rm s}, 
\delta_0^\fifty=-28^\circ~55'~00.445"$ (Lane 1979).
Converting the B1950 coordinates to J2000 coordinates, we obtain 
$\alpha_p^\twoth=12^{\rm h}~51^{\rm m}~26.282^{\rm s}, \delta_p^\twoth=+27^\circ~07~42.01"$
and
$\alpha_0^\twoth=17^{\rm h}~45^{\rm m}~37.224^{\rm s}, \delta_0^\twoth=-28^\circ~56'~10.23"$.
In order to carry the IAU definition of the Galactic plane forward to 
J2000 coordinates, we need to determine a new value of $\theta$.
We do this by requiring that the J2000 coordinates of the origin
gives zero longitude, which yields $\theta^\twoth=122.932^\circ$.
Note that at the position of the Galactic center, the projection of the
Galactic plane is at a position angle of $31.72^\circ$ and $31.40^\circ$ 
east of north in B1950 and J2000 coordinates, respectively. 
\citet{Blaauw60} suggest a probable error of $\pm0.1$~degrees in the
orientation of the Galactic pole, and hence the Galactic plane.

\clearpage

\end{document}